\documentclass[12pt,letterpaper]{article}
\usepackage{jheppub}
\usepackage{graphicx}
\input{epsf}
\usepackage{epsfig}
\usepackage{epstopdf}
\usepackage{enumerate}
\usepackage{calligra}
\usepackage{subfigure}
\usepackage{color}
\usepackage[normalem]{ulem}

\definecolor{purple}{rgb}{0.6, 0, 1}

\newtheorem{definition}{Definition}
\newcommand{\be}{\begin{equation}}
\newcommand{\ee}{\end{equation}}
\newcommand{\ba}{\begin{array}}
\newcommand{\ea}{\end{array}}
\newcommand{\bea}{\begin{eqnarray}}
\newcommand{\eea}{\end{eqnarray}}

\def\IZ{\mathbb{Z}}

\def\CO {{\cal O}}

\def\CCC {{\cal C}}

\def\CA{{\cal A}}

\newcommand{\eq}[1]{Eq.~(\ref{eq:#1})}

\def\one{{\hbox{ 1\kern-.8mm l}}}

\def\ba{\bar{a}}

\def\vev#1{\langle{#1}\rangle}

\def\boldclass{\bf\sf}
\def\P{{\boldclass P}}
\def\NP{{\boldclass NP}}

\def\MA{{\boldclass MA}}

\def\BPP{{\boldclass BPP}}

\def\BQP{{\boldclass BQP}}
\def\QMA{{\boldclass QMA}}

\def\searchS{{\bf S}}
\def\searchH{{\bf H}}
\def\searchR{{\bf R}}
\def\searchI{{\bf I}}

\def\itclass{\it}

\def\CC{{\itclass CC}}

\def\SIMCC{{\itclass SIM-CC}}
\title{Computational complexity of the landscape II
--\ Cosmological considerations}

\author[1]{Frederik Denef,}
\author[2]{Michael R. Douglas,}
\author[1]{Brian Greene,} 
\author[1]{and Claire Zukowski}

\affiliation[1]{Department of Physics, Columbia University, 538 West 120th Street, New York, NY 10027} 

\affiliation[2]{Simons Center for Geometry and Physics, Stony Brook University, Stony Brook NY 11794} 
 
\emailAdd{fmd7@columbia.edu}
\emailAdd{mdouglas@scgp.stonybrook.edu}
\emailAdd{bg111@columbia.edu}
\emailAdd{cez2103@columbia.edu} 

\abstract{We propose a new approach for multiverse analysis based on computational complexity,
which leads to a new family of ``computational'' measure factors.
By defining a cosmology as a space-time containing a vacuum with specified properties (for example small cosmological constant) together with rules for how time evolution will produce the vacuum, we can associate global time in a multiverse with clock time on a supercomputer which simulates it. We argue for a principle of ``limited computational complexity" governing early universe dynamics as simulated by this supercomputer, which translates to a global measure for regulating the infinities of eternal inflation.  The rules for time evolution can be thought of as a search algorithm, whose details should be
constrained by a stronger principle of ``minimal computational complexity.'' 
Unlike previously studied global measures, ours avoids standard equilibrium considerations and the well-known problems of Boltzmann Brains and the youngness paradox. We also give various definitions of the computational complexity of a cosmology, and argue that there are only a few natural complexity classes.
}

\begin{document}
\maketitle

\section{Introduction}

Since the advent of the string landscape, there have been two contrasting perspectives on string/M theory. The first focuses on the enormous range of vacua, and a similar enormous range of low energy dynamics, and concludes that string theory predicts ``everything'' -- and hence nothing. This perspective argues that a theory which can accommodate any and all measurements is a theory without predictive content. The second  perspective focuses on the fact that researchers have yet to produce a first principles string model that agrees with all precision particle physics and cosmological measurements. This perspective highlights the absence of even a single, explicit string theoretic model that would definitively establish the theory's capacity to embrace known physics.

While the two perspectives are not in contradiction, they provide curious extremes. At one and the same time they suggest that string theory is so flexible that it can trivially accommodate any observations, and yet so intractable that it has yet to accommodate the known observations.  How can this be?  Is it simply a consequence of our limited understanding of
and ability to work with string theory, a situation which will improve with time?  Or is there some actual barrier to
constructing explicit string theoretical models that embrace precision physics?

About a decade ago \cite{Denef:2006ad} this question motivated two of us (FD and MRD) to bring complexity theory to bear on the question of locating models in the string landscape that meet particular phenomenological requirements, most notably having a cosmological constant on par with observations. We found that in the simplified
models often now considered, the problem is NP hard, giving insight into the paucity of explicit models that fit the data. 

In this paper, we continue this line of research but embed it in a cosmological framework. It has long been realized that cosmology is the natural context for multiverse theorizing. Cosmology elevates the abstract notion of possible universes into the instantiated realization of actual universes, ones that are created by dynamical mechanisms and which populate a multiverse. In this setting, the question we ask is this: If a theorist searching mathematically for our universe in the string landscape is tackling such a difficult problem, how does the universe end run this obstacle and yield our vacuum cosmologically? 

The standard framework for answering this question, an approach on which researchers have for some time pinned their hopes, is that we will one day acquire a measure on the multiverse that allows us to assign the moniker ``likely" or ``typical" to particular universes with certain specific, detailed properties, and that, assuming the program succeeds, our universe will be one of these. With no judgment intended, because this approach hinges on our universe being ordinary, it has sometimes been called the principle of mediocrity \cite{Vilenkin:2011dd}. In short, nature finds our vacuum in the multiverse because, at least among those that can support life, universes like ours are a dime-a-dozen.

While we won't give a comprehensive history of efforts to rigorously realize this approach, suffice it to say that it has proven challenging. These difficulties have inspired us in this paper to propose a different approach to these cosmological questions. Rather than  seeking to pinpoint run-of-the-mill vacua in an infinite multiverse, we instead draw attention to vacua that are easily produced through cosmological dynamics. Roughly speaking, rather than seeking the equilibrium distribution of universes in an aged multiverse, we consider anthropically hospitable\footnote{
By ``anthropically hospitable'' (or just hospitable)
we mean a universe in which the low energy physics allows for the formation
of matter and structure, some sort of nontrivial chemistry, and whatever else is needed for the evolution of observers,
but not necessarily containing observers.  We will explain the distinction below, 
but the main point is that it should
be easy to check the condition of hospitability given the laws of low energy physics.}
universes produced in a youthful multiverse. The conceit, then, is that nature found us not because our type of vacuum is ubiquitous but rather because our type of vacuum is easily generated through cosmological evolution and hence appears early on in the unfolding of the multiverse.

While this idea seems intuitively plausible, so far as we know the present work is the first to try to make it precise, and the first to suggest any arguments based on it which could lead to physically testable consequences.  How would one make it precise?  Our goal will be the same as in many previous
discussions of multiverse cosmology:
a measure factor, {\it i.e.} a probability distribution over hospitable universes.
As in those works, one hopes to use the measure factor to compute the probability of specific testable
features of the hospitable vacua, say superpartners
with mass below some scale, or constraints on cosmological parameters, to
get testable predictions.
Furthermore we will start out in the same way as previous works, by making
three postulates: the laws of physics, the initial conditions, and a definition of cosmological time.
In \S \ref{s:measures}, we briefly review the existing framework for measure factors in eternal inflation,
following particularly \cite{Garriga:2005av}, as our point of departure. 

Unlike previous discussions, our primary definition of cosmological time will {\it not} be
physical time within the multiverse, but rather a measure of the difficulty of creating a particular
universe (or region of space-time) within the multiverse.  With such a definition, the proposal is
simple. Given a particular history for the multiverse, a unique hospitable vacuum is selected:
the {\it earliest} one, meaning the vacuum with the smallest cosmological time $t$ by which it 
is created and checked to be hospitable.  While this may sound as if it will single out a specific
preferred vacuum, it will not.  As we review later, the cosmological dynamics of the 
multiverse is not deterministic but rather probabilistic and/or quantum mechanical, and this 
dynamics will
lead to a probability distribution over histories.  Following the rule we just gave, this
probability distribution will determine a probability distribution over hospitable vacua, the measure
factor.

Thus, we must explain what we mean by ``the difficulty of creating a particular
universe within the multiverse,'' and we do this in \S \ref{sec:supercomputer}.
Here is the main novelty of our discussion: By ``difficulty of X'' we
will mean the computational cost of realizing X within a {\it simulation} of the multiverse, carried
out by a hypothetical quantum supercomputer which can simulate the laws of physics.
By the hypothesis of universality of computation (the Church-Turing thesis, and its quantum 
generalization), if the laws of physics can be made precise, their predictions can be reproduced
by computations performed by a computer.  Thus we imagine a supercomputer built out of quantum gates, 
and define the difficulty of realizing X as the number of quantum gates needed to implement a time
evolution which starts with a state encoding the initial conditions and produces a state containing
X.  This idea will also allow us to define the other terms in our definition above, such as ``checking
that a vacuum is hospitable.''  We will think of the supercomputer as carrying out an explicit
search for a hospitable vacuum, and the total number of quantum gate operations required to
find the vacuum is the ``difficulty'' or ``cosmological time.''  In this way we will render precise the
hypothesis that ``our type of vacuum is easily generated through cosmological evolution."

Although we do not yet have a formulation of string theory or quantum gravity which can be
simulated in this way, one might expect that a simple question such as ``how many quantum
gates are needed to simulate physics in a given region of space-time'' would have a simple answer.
One conjecture (which probably dates back to very early discussions of quantum computing)
is that the number of gates needed to simulate a region $R$
is proportional to its volume in Planck units.
Since \cite{Denef:2006ad}, the topic of space-time geometry and computational complexity theory has received more study, and this allows us to make a better conjecture.  Following \cite{Brown:2015bva}, we
conjecture that the number of quantum gates needed to simulate a region $R$ is proportional to
the Einstein-Hilbert (or full string theory) action.  This conjecture will motivate a definition of 
``action time'' in \S \ref{ss:act1}, which we study in detail in \S \ref{sec:ActionTime}.

Now, given a measure of the computational cost of simulating a region of space-time, one can propose
and study various cosmological search algorithms, which try to find hospitable vacua at minimal cost.
The simplest search algorithm is one which simply ``watches'' the evolution of the multiverse along a
geodesic and takes the first hospitable vacuum it sees.  This algorithm is a simple variation on the proposal of
\cite{Garriga:2012bc}, which is the previous proposal closest to the one we make here.
However even this simple variation can lead to major changes in the resulting measure
factor, as we will show in \S \ref{s:toymodel}.

More generally, we make a definition which encompasses a wide range of search algorithms, which can make choices among the various alternatives available for future exploration, based on the results of previous observations.
This generality is a key difference between our proposal and earlier work.  

Even very simple search algorithms can lead to totally different search processes and measure factors
than those found in previous work.  As another example, we consider a search process which avoids spending
time simulating very long-lived vacua, essentially by imposing a ``cutoff'' on the amount of time
it will spend in any branch of the search.  Eventually, the supercomputer will abandon simulating a long-lived
vacuum, in favor of simulating other vacua.  Since vacuum lifetimes in
realistic landscapes can be extremely long (given by double exponentials), the cutoff
time can be taken very large and its precise value need not be important, but this change explicitly excludes
the dominant vacua of the standard discussions, as well as eliminating paradoxes such as Boltzmann Brains.

A complete derivation of a measure factor along these lines will depend on many additional choices.
Beyond specifying evident physical features, including initial conditions and the definition of ``hospitable,'' 
new choices enter: details of the definition of the supercomputer and specifics of the program it is running.  
Ultimately the proposal is interesting to the extent that its predictions are independent
of such choices.  Much of our subsequent discussion will address this point, arguing that any measure
factor of this type will differ significantly from those of previous discussions, that its general form
need not depend strongly on details of the initial conditions and particular search procedure, and that
the postulate of minimizing the computational cost of the search will drastically constrain the choice
of algorithm.

On the other hand, some aspects of the landscape, such as the general nature of the initial conditions, and
the general structure of tunneling rates, are crucial to making any predictions.  
These aspects clearly depend on the detailed definition of the landscape in string compactifications (or otherwise)
and understanding them would be important goals for future research. 

Finally, we make some comments about a more abstract version of this discussion, which 
defines the complexity class of a cosmology.  Our proposal was inspired by
computational complexity theory, and particularly the idea of computational reduction.
Can we give meaning to questions such as ``is the problem of finding a vacuum with small
cosmological constant in \P, \NP\ or some larger complexity class?" 

\section{Measure factors in semiclassical gravity}
\label{s:measures}

Let us begin by briefly summarizing the theory of measure factors in eternal inflation,
mostly following the discussion in \cite{Garriga:2005av}.  Our discussion will be rather sketchy and readers new to this topic should consult an introduction such as \cite{Guth:2000ka,Guth:2007ng}, or one of the many reviews such as \cite{Freivogel:2011eg}. We also focus here only on the formalism and defer any more physical discussion to \S \ref{s:toymodel} and \S \ref{s:measureresults}.

For definiteness we take as
our landscape a configuration space parameterized by scalar fields $\vec\phi$ with
a potential $V(\vec\phi)$.  The local minima of $V(\vec\phi)$ are vacua, and we will label
each vacuum with a distinct index $i=1,2,\ldots$.  We assume the anthropic solution
of the cosmological constant problem, according to which cosmological dynamics will produce
a large number of diverse vacua, of which a few will satisfy anthropic constraints including
small cosmological constant, and that these few will be post-selected as the only ones
which could possibly be observed.  Thus our first goal is to identify cosmological
dynamics which, starting from generic initial conditions, will produce space-time histories
containing large numbers of vacua of the various types.  We need to propose a definite
way to count the number of vacua at time $t$, leading to a variable $N_i(t)$ for each type.  
We furthermore need a definition of the expected number of observers which will be created
in vacuum type $i$ (perhaps depending on time), call this $X_i(t)$.
We can restrict attention to the anthropically selected vacua (those with $X_i(t)>0$),
call this set $\CA$. 
Then we assert that the probability $P_i(t)$ that an observer will exist in vacuum $i$ is the
fraction
\be \label{eq:defmeasure}
P_i(t)=\frac{ N_i(t)\, X_i(t)} {\sum_{ j \in \CA } N_j(t)\, X_i(t)} ~.
\ee

The ``principle of mediocrity'' then tells us that 
we should live in a vacuum with large probability according to this measure.  
Rather than try to make the word ``large'' precise, we can instead
use the measure to define probabilities of observables.  As an example, consider the scale of supersymmetry
breaking $M_{susy}$ (in a low c.c. vacuum this is $\sqrt{3}$ times the gravitino mass).
The above discussion implies that its observed probability distribution will be
\be \label{eq:msusy}
P(M_{susy} < M) = \sum_{i \in \CA, M_{susy,i} < M} P_i .
\ee
If it turns out that the value of this observable in our universe
is low probability, that is evidence against the theory.

The importance of the factors $X_i$ depends on whether the observable under consideration is correlated with the $X_i$.  At one extreme of maximal correlation, 
we have the observable ``expected number of observers.''  Many interesting cosmological observables,
such as the cosmological constant itself and the size of density fluctuations, also have a direct
influence on the $X_i$'s. To compute the distribution of these observables, it is important to know
this effect.

At the other extreme, there are believed to be many observables with essentially no correlation
with the $X_i$.  A standard example is the $\theta_{QCD}$, which governs one form of CP violation
and is small in our universe.  Because we believe $\theta_{QCD}$ is uncorrelated with the $X_i$,
we believe that the actual distribution of $N_i$'s is peaked on small $\theta_{QCD}$, so that this
prediction will hold.
Another arguable example is $M_{susy}$, conditioned on this being well above the
scales of nuclear and atomic physics (as in our universe).  It is not apparent that there should be
any relation between this observable and the existence of observers.  If not, and if the number of
vacua is sufficiently large that accidental correlations average out, then we would not need to know the $X_i$'s to compute \eq{msusy}, and information about $M_{susy}$ in our universe would give evidence about the
dependence of the $N_i$'s on $M_{susy}$.

In the rest of this section we outline how the numbers $N_i(t)$ are determined in the standard
discussions, saving all discussion of the factors $X_i(t)$ for
\S \ref{sec:youngness} when we explain the analogous point in our proposal.

Let us assume that the initial conditions have constant $\vec\phi$ in some region
set to be a generic point in configuration
space with $V(\vec\phi)>0$ and $V'(\vec\phi)\ne 0$.
We recall that in classical general relativity, regions of space-time
with $V(\vec\phi)>0$ undergo inflation, an exponential expansion locally modeled by
a de Sitter space-time with cosmological constant $\Lambda=V(\vec\phi)$, and thus with
expansion rate $H^2=8\pi V(\vec\phi)/3$.
The dynamics for $\vec\phi$ will make it decrease with time, and eventually the universe will
reach a steady state with $V'(\vec\phi)=0$, one of the many vacua.
In this simplification of the dynamics, there is no mechanism for creating a multiverse.

Eternal inflation
\cite{Steinhardt:1982kg,Linde:1982ur,Linde:1982gg,Vilenkin:1983xq,Linde:1986fc,Linde:1993xx} (for a brief
review of the basics see e.g.\ \cite{Guth:2000ka}) occurs more or less
naturally in any landscape model once
quantum mechanics is taken into account. Even given our earlier assumption
that $\phi$ was constant over some initial spatial slice, quantum
fluctuations of various kinds will spoil this uniformity.  
As a consequence, the different causal regions created by inflation
can have different $\phi$ values.  

If the expected number of causal regions after one Hubble time with
$H_{final} \ge H_{initial}$ is greater than one, then on average the number of
inflating regions will grow with time, and with high probability
inflating regions will always exist, hence the name.  On the other
hand, it can be shown that any given world-line is overwhelmingly likely
to eventually enter a conventional slow roll regime.  Thus, this
provides a mechanism to populate all of configuration space and
create a multiverse, independent of the choice of initial conditions.
Indeed, the general expectation is that since volume factors are so large,
all memory of the initial conditions will quickly be lost, perhaps
leading to a universal probability distribution.

Another model for quantum fluctuations is to consider
events in which the $\phi$ field can tunnel through a potential
barrier, as
is familiar in quantum mechanics.  In field theory, such
processes are described by instantons, interpolating solutions of the
Euclidean equations of motion.  The original example in semiclassical
quantum gravity is the Coleman-de Luccia instanton
\cite{Coleman:1980aw}.  This describes an event in which a small
bubble of new vacuum is nucleated inside the old vacuum, separated by
a domain wall in which the scalars interpolate between the two
critical points.  It is followed by a classical expansion of the
bubble wall, which quickly approaches the speed of light.

One can compute the probability for such a transition from the Coleman-de Luccia instanton action in the thin-wall approximation. For tunnelings to vacua
of lower cosmological constant, {\it i.e.} downward tunneling, its general form is 
\bea \label{eq:cdlrate}
\Gamma_{i,j} &=& A_{i,j} e^{-B_{ij}}~. \\
B_{ij} &=& \frac{27\pi^2}{2} \frac{\tau^4}{M_P^2 |\Delta \Lambda|^3} r(\tau, \Delta \Lambda)~,
\eea
where $M_P$ is the 4D Planck mass, $\Delta \Lambda=\Lambda_i-\Lambda_j$ and $A_{i,j}$ is an order
one factor.  The function $r$ is important in detailed considerations as it can modify the powers of
$\tau$ and $\Delta\Lambda$ in the exponent, but we will not need it here.
The upward tunneling rate is 
\be \label{eq:upward}
\Gamma_{j,i} = \Gamma_{i,j} \exp \left[24\pi^2 M_P^4 \left(\frac{1}{\Lambda_j}-\frac{1}{\Lambda_i}\right)\right]
\ee
which is also determined by detailed balance.

While in the large c.c. regime $\Lambda \sim M_P^4$, these tunnelings are relatively unsuppressed,
once $\Lambda \ll M_P^4$ they are generally quite rare (this depends on the bubble tension of course).

Because of inflation, this process does not destroy the old universe,
but instead should be thought of as creating a new causal region.
A bubble never overtakes the entire space, since
observers further than a Hubble radius $1/H$ from the point where
the bubble nucleated inflate away too fast for even light to catch
up. This can be seen as follows. Light rays travel on paths along
which $ds=0$, i.e.\ $d\vec{x}/dt = e^{-H t} \vec{u}$, with
$\vec{u}^2=1$. Integrating this gives
\be
 \vec{x}(t) = \vec{x}_0 + \frac{1}{H}(e^{-H t_0} - e^{- H t}) \vec{u}~.
\ee
Therefore the longest coordinate distance a light ray, and therefore
any signal, can travel is $|\Delta \vec{x}|=e^{-H t_0}/H$. To get
the physical metric distance we have to
multiply by $e^{H t_0}$, which gives a distance $1/H$. Thus, an
observer can only be affected by events within a sphere of radius
$1/H$, which is this observer's ``event horizon''. For the same
reason, most bubbles will not coalesce, as they are inflating away
from each other too rapidly.

Inside this bubble, new bubbles can form, and so on, \emph{ad
infinitum}.  Moreover, although transitions to lower cosmological
constant are preferred, in a space with $\Lambda>0$, transitions to
higher cosmological constant are possible as well.  
The resulting process produces a multiverse very similar to that of the eternal
inflation process we discussed earlier.  Indeed, in \cite{Garriga:2005av} 
it was argued that the two processes are limiting descriptions of the same dynamics,
and can be described together within the same formalism.  In this work we will 
generally use the Coleman-de Luccia tunneling language to describe the transitions 
between vacua which populate the multiverse.

The interpretation we just discussed makes sense if both the initial and final vacua
have positive cosmological constant, {\it i.e.} are approximately de Sitter.  The string
landscape also contains Minkowski and anti-de Sitter vacua and their role in 
this discussion is not fully understood.  One can argue -- very convincingly for Minkowski
and less so for anti-de Sitter -- that they are local end points in the dynamics, which
do not tunnel back to de Sitter vacua.  This can be modeled in equation \eq{markov}
by setting all transition rates out of such vacua to zero.  Alternatively, as conjectured in
\cite{Garriga:2012bc}, it may be that AdS
vacua are not terminal, but instead ``bounce'' to de Sitter vacua, in some way which can be computed in the underlying fundamental theory.  In this case one could treat the AdS vacua on the same footing as the dS vacua,
with the difference that the rates $\Gamma_{j,i}$ for transitions from AdS to dS are not given by
\eq{upward} but instead by some yet-to-be discovered formula.  An alternative would be to 
``integrate out'' the AdS vacua and instead add terms describing transitions with intermediate
AdS vacua to the dS to dS rates.  This would be simpler to the extent that AdS vacua are short lived (so
are clearly not hospitable) and to the extent that AdS to dS rates do not depend on the particular AdS intermediary.

The upshot of the discussion so far is that early cosmology contains a stochastic dynamics
which can populate the various vacua in a probabilistic way.  The next step is to solve for
the dynamical evolution of this population of vacua.

\subsection{Markov process}\label{sec:markov}

In \cite{Garriga:2005av} and many other works, a formalism was developed
to model these physical processes and derive a measure factor.   Let us state their basic results.
The dynamics of eternal inflation will be described as a Markov process
where the state is a vector $f^i(t)$ whose elements are the fractional co-moving volume 
in vacuum type $i$ at time $t$.\footnote{We will sometimes refer to this 
imprecisely as the ``number of vacua of type $i$.''}  
The dynamics is then
\be \label{eq:markov}
\frac{\partial f^j}{\partial t} = \sum_j M^j_{\ i} \; f^i(t)
\ee
where
\be \label{eq:defM}
M^j_{\ i} = \kappa_{j,i} - \sum_k \kappa_{k,i} \delta^j_{\ i}
\ee
is a sum of a matrix $\kappa_{j,i}$ of the rates for transitions from the current vacuum type
$i$ into new vacuum types $j$, and a term for transitions out of the current vacuum type.

For a pair of de Sitter vacua $i$ and $j$, the matrix $\kappa_{i,j}$ is 
\be \label{eq:defkappa}
\kappa_{i,j} = \Gamma_{i,j} \frac{4\pi}{3} H_i^{\beta-4}
\ee
where $\Gamma_{i,j}$ is the tunneling rate as in \eq{cdlrate},
$H_i$ is the expansion rate in the vacuum type $i$, and the power $\beta$
depends on the precise definition of global time $t$ (see the next subsection).

Given the assumption of the previous subsection, the matrix $\kappa_{j,i}$ for
$i$ a Minkowski or AdS vacuum is taken as zero, while for the case of $i$ de Sitter
and $j$ Minkowski or AdS we take the probability as defined by the Coleman-de Luccia
tunneling process.   In the probability literature, states with zero outgoing transition
rates are called ``terminal states,'' and thus 
the Minkowski and AdS vacua are referred to as ``terminal vacua.''
The existence of terminal states
has many important consequences for the solutions of \eq{markov}.
For example, in the absence of terminal vacua, the smallest magnitude eigenvalue of $M$ will be $q=0$,
while given terminal vacua it can take values less than zero.  We will return to this below.

Given a knowledge of the set of vacua with their energies and tunneling rates, because 
\eq{markov} is a linear equation, it is (at least conceptually) easy to solve, by finding
the eigenvectors and eigenvalues of the matrix $M$.  One can then use the resulting
 $f_i(t)$ in \eq{defmeasure} to derive a measure factor, at least as a function of time.
The linearity is to be contrasted with rate
equations in chemistry, nuclear physics and other disciplines which are generally nonlinear.
This difference is a good reason to think that an {\it ab initio} approach is more promising
for early cosmology than it would be for most physical systems.

\subsection{Global time}

We have skipped over (and will not discuss in detail) the most subtle part of this formalism, which
is the definition of time.  Given a cosmological history in space-time, to define the fractional co-moving volumes $f_i(t)$, one must define the time $t$.  The original and conceptually simplest way to
do this is to pick a global time coordinate $t$ on space-time and count vacua on each hypersurface
of fixed $t$.  Many other approaches have been developed and we refer to \cite{Freivogel:2011eg} for
a recent review, but we will not use them here.

Conceptually the simplest choice of time coordinate is proper time.
To define it, we choose an initial hypersurface $\Sigma_0$ 
on which to set $\tau=0$, and a future oriented time-like vector
field on this hypersurface.  This data defines a family of time-like geodesics, and the time $\tau(p)$
for a point $p$ is the proper time at which one of these geodesics reaches it.

Given proper time, one can define a one-parameter family of related time coordinates, as
\be \label{eq:defbetatime}
dt_\beta = [H(\phi(\tau))]^{1-\beta} d\tau~.
\ee
The choice $\beta=1$ is proper time, while the choice $\beta=0$ amounts to using the
logarithm of the scale factor as time and is thus called `scale factor' time.\footnote{Note that a standard equilibrium cutoff with the choice $\beta=0$ was shown to avoid the Boltzmann brain problem~\cite{DeSimone:2008if}.}
Later we will find that $\beta=2$ appears naturally in our discussion, and 
at that point $t_{\beta=2}$ will receive the name ``action time.''
Note however that the $\beta$-dependent factor in \eq{defkappa}
(a power of $\Lambda_i$) is usually subdominant
to the extremely small (often double exponential) factors in the decay rates $\Gamma_{i,j}$.
Thus we can to some extent continue the discussion without fixing $\beta$.  Of course
there are other definitions of global time besides \eq{defbetatime} and the derivation
of \eq{defkappa} would have to be reconsidered for these.  In particular our considerations
in \S \ref{sec:supercomputer} would produce such definitions, but we will leave the question of  
more carefully generalizing \eq{defkappa} to future work.

Given a definition of global time $t$ and thus of the hypersurfaces at each global time,
one can define $f_i(t)$ in terms of the volume of the space-like hypersurface for which the
fields $\vec\phi$ are close to the minimum of $V$ associated with the vacuum $i$.
This is conceptually straightforward if the total volume of the hypersurface is finite, but
fraught with difficulty if it is not.  Since the total volume will grow exponentially, even
if we grant that the initial volume is finite, any considerations at large $t$ -- and especially
any attempt to take the limit $t$ to infinity -- will be very subtle. (See, for instance, \cite{Guth:2007ng}.)

\subsection{Initial conditions}

So far the discussion allows for arbitrary initial conditions, encoded in the vector $f^i(t=0)$.
Now there have been proposals for preferred initial conditions, most famously that of Hartle
and Hawking \cite{Hartle:1983ai}.  Working any of them out in detail
requires knowing more about the configuration space.  Indeed it is likely that the simplification
we followed of reducing this to a set of vacua labeled by an index $i$ is inappropriate for
understanding the initial conditions.  One might need to use a wave function or probability
density defined on the space of scalar fields $\vec\phi$, perhaps augmented by choices for the topology,
fluxes and other structures in the extra dimensions. It is hard to seriously discuss this without
better knowledge of string theory (or a hypothetical different theory of quantum gravity).

In the theory of observable inflation (as is used in studying the period of inflation which is hypothesized
to lead to predictions for structure, for the cosmic microwave background, and otherwise), an important
general principle is the independence of predictions from many details of the initial conditions -- one
says they are ``inflated away.''  An analogous principle in the theory of eternal inflation is that the
measure factor should not depend on details of the initial conditions, $f^i(0)$.  One can find support
for this claim in the mathematics of the Markov process \eq{markov} and the idea of ``mixing time.''
As is familiar, the solution of \eq{markov} is most easily expressed by making a linear transformation
on the state space to a basis of eigenvectors, in which $M$ is diagonal.  On general grounds, one then
expects that the smallest absolute magnitude eigenvalue is non-degenerate, and in this case the infinite time limit of
$f^i(t)$ is almost always proportional to the corresponding eigenvector (since it will almost always
have nonzero overlap with the initial conditions), often called the dominant eigenvector.
The time taken to approach this limit is determined
by the mixing time, which is the gap between the smallest and second-smallest absolute magnitude eigenvalue.  Given
that this is nonzero (assuming a non-degenerate spectrum and, in the case of an infinite number of vacua, the absence of
a continuous part), there will be some
time $T$ after which the distribution has, to any required accuracy, converged
to a universal value.  This is of course the standard discussion of the approach to equilibrium in
statistical physics, and what we have just argued is that it is reasonable to believe that cosmological dynamics has an equilibrium.

In most discussions of the measure factors derived from eternal inflation, it is assumed that 
cosmological dynamics has had sufficient time to reach equilibrium.  Thus the state $f^i$
becomes essentially equal to the dominant eigenvector of $M$.  The consequences of this 
were analyzed in \cite{Garriga:2005av} in general and in \cite{SchwartzPerlov:2006hi} for a
Bousso-Polchinski landscape.  By arguments we will review later, the dominant eigenvector 
will have most of its support on the metastable vacuum with the longest lifetime.
Assuming this vacuum is itself not hospitable, the relevant measure factor is the 
vector of tunneling rates from this dominant vacuum to the hospitable
ones.

The upshot is that the theory of measure factors for eternal inflation combined with
some natural assumptions, most notably that cosmology reaches its equilibrium, leads 
to a fairly specific prescription which can be applied to the string landscape.  We will
discuss some conjectured properties of the results in \S \ref{s:measureresults}, following \cite{Douglas:2012bu}.

\section{The cosmic supercomputer}\label{sec:supercomputer}

The core idea of the present work is that our universe should be ``easy to find'' within the multiverse,
and global time is meant to quantify this.  To make this precise, we need to say what we mean
by finding hospitable vacua, and express the process as a specific series of steps which can be enumerated,
so that the number of such steps becomes a measure of time.  In other words, we want a computational
description of cosmology allowing us to
precisely delineate the hypothesis that early cosmology is computationally limited.

Let us say this in a more colorful way.
Suppose we have a quantum supercomputer that can simulate the laws of physics -- for the
purposes of this paper, string/M theory -- throughout the entire universe.  Suppose it can execute
the computations required to simulate the entire 14 billion year history of the observable universe in, say, 
a single second of our subjective time.

How many computations are needed?  The basic element of a quantum computer is a quantum gate,
which makes a unitary transformation on the state of some small number of qubits.
One can reasonably suppose that the number should be related to the length of time and amount of
space being simulated.  By modestly generalizing an interesting recent observation
\cite{Brown:2015bva}, we conjecture
that simulating the observed universe will take about $10^{120}$ quantum gate operations.
Call this number $N_Q$.  While large, it is finite, and it is the natural ``benchmark'' with
which to compare other computations we might postulate as relevant to cosmology and our universe.

Now according to string/M theory, our universe has six or seven
extra dimensions, curled up in some topological manifold, adorned with branes, fluxes and the like.
Suppose we do not know the details of this additional structure, so we ask our supercomputer to find it and then carry on with the simulation. Without foreknowledge of the extra dimensional structure, 
all we can ask is that the supercomputer find candidate geometrical data which are able to reproduce all experimental and observational results to date.
Since this task is all the more difficult, we acquire a yet faster supercomputer.  If it takes $N_Q$ quantum operations to simulate our universe with a known choice of vacuum, we acquire one able to carry out $2N_Q$ quantum operations in a second, or even some
large constant $k$ times $N_Q$ (but with $k\ll N_Q$).    Could such a supercomputer find a candidate realistic vacuum and simulate a universe like ours?

The answer is not obvious, and in the following we will argue that the answer might well be ``no.''  
More pointedly, we will argue that the answer is ``no''
if our supercomputer simulates the dominant physical paradigm we currently use
in quantum cosmology: eternal inflation and the multiverse, as we discussed in \S \ref{s:measures}.

On the other hand, if the supercomputer follows a more efficient strategy, 
then as we will explain, the answer might well be ``yes.''
Furthermore, the solution (the structure, including adornment, of the extra dimensions)
which the supercomputer finds using the more efficient strategy will likely be different from that identified by 
the standard strategy (of measure factors derived from eternal inflation)
thus leading to distinct predictions for yet-to-be discovered physics.

In theoretical computer science, one generalizes such questions further still.  If we have a class
of problems with varying size $N$, one can ask if they are each solvable given $k N^s$ operations for
some particular fixed $k$ and $s$.  If so, one says that these problems live in the complexity class \P.
Can our supercomputer find a realistic vacuum using $k N_Q^s$ operations for some $k$ and $s$?

Of course if the number of operations needed to find a realistic vacuum
is finite, the answer to this question is trivially ``yes.''  Defining a complexity
class always requires defining a class of problems, so one cannot literally ask whether
the problem of realizing our specific universe is in \P\ or not.  We discuss a version
of this question which postulates a class of problems in \S \ref{s:compclass}.

\subsection{Rules for simulation}
\label{ss:globaltime}

Here we begin to make precise the proposal we've outlined.
Let us ask whether a computer with specified power can
{\it simulate} early cosmology. By that we mean the computer can carry out computations that yield
results of measurements which a hypothetical observer, with access to large parts of the
multiverse, could obtain, and then uses
the results to find (in a sense we will make precise shortly) a hospitable vacuum.

We will always enforce the principle that an observer in a single universe can not directly discern
that there is an underlying simulation. Thus, for example, when the supercomputer
makes observations of a given universe, it is not allowed to use the results to
affect any observation subsequently made by an observer confined to that universe.  
At first one may wonder whether anything at all can be done under this constraint, but we will shortly 
explain how it can.  Thus, to make these arguments, 
we need not take any position on whether our universe ``actually is'' a simulation. Yet, this ``principle of unobservability'' notwithstanding,
we can derive testable consequences of the sort explained in the introduction -- namely,
that certain types of hospitable universes are easier to find and thus are preferred candidates
for the reality we inhabit.  We return 
to comment on this and other philosophical points in the conclusions.

To simulate quantum cosmology, we need a quantum computer.  We follow the textbook definition
of a quantum computer as a quantum system in which the
wavefunction is a tensor product of two-state systems called qubits, and 
in which the computation is implemented by acting with a unitary operator.
Not having a complete formulation of quantum gravity, we cannot give any details about how
one would use such a computer to simulate the laws of physics.  But let us grant that by doing a simple 
measurement on its wave function, one can test the proposition that
``the multiverse contains a vacuum of type $i$,'' or perhaps find the expected number
$\vev{N_i}$ of universes of type $i$.  Thinking about this
suggests a natural definition of global time: the global time $t(\CO)$ of an observation
$\CO$ made in a cosmology,
is the minimal number of quantum operations that a quantum computer simulating this
cosmology (from a specified initial condition) would need to perform the observation.
Later we will express our proposal in these quantum terms, but for now let us consider the semiclassical
limit.

In the semiclassical discussion, the state is defined on a space-like surface, and consists of
a metric tensor, its canonical momentum (the extrinsic curvature), and canonical variables
for the other fields in the theory.  Time evolution is not deterministic, but probabilistic.
We will think about it using the following approximation: with probability $\sim 1-e^{-S/\hbar}$
the time evolution is deterministic and generated by a classical Hamiltonian, 
but tunneling processes can also take place with probability $e^{-S/\hbar}$.  While these can 
usually be associated with instantons (imaginary time solutions of equations of motion), we will
not use this but instead think of tunneling events as instantaneous jumps in phase space which change
the configuration locally, and which occur with a probability which can, in principle, be inferred from
the local configurations before and after the event. 

Unlike other field theories, in gravity the Hamiltonian is gauge dependent, and depends on the
choice of time coordinate.  An infinitesimal time evolution is defined by a choice of lapse 
function and shift vector, and it can advance the space-like surface by a time-like vector field
which depends on spatial position in an arbitrary way.  This is the point in the discussion at which 
we will need to make a choice that will lead to a preferred definition of global time.

To do so, imagine that the state of the underlying supercomputer includes the
physical state on a hypersurface $\Sigma = \Sigma(0)$, as we just discussed, and perhaps internal state
variables in some
simple state -- in particular, one which does not contain detailed information about the landscape.

Then, the basic choice that the supercomputer can make, consistent with the laws of semiclassical
gravity, is how to advance the hypersurface $\Sigma$.  In the language of the canonical formalism,
it can choose a lapse and shift vector, multiply these by the Hamiltonian and momentum densities,
and generate the resulting time evolution, thereby simulating a region of space-time with past
boundary $\Sigma$ and future boundary some new $\Sigma'$.\footnote{In quantum gravity,
the Hamiltonian and momentum operators are constraints, so one must phrase this more carefully.
In quantum cosmology one conditions the wave function on the scale factor, and reinterprets
the scale factor dependence as time, see {\it e.g.} \cite{Halliwell:1990uy}.  Using WKB this can be reduced to the
canonical formalism in the semiclassical limit.  We will speculate about how to make a fully quantum discussion in the conclusions. }
Our proposal is to allow this choice to depend on observations made by the
supercomputer -- specifically the types of previously created
universes, tunneling events and lifetimes -- by some specified search algorithm.  Thus, as a
particular instance of the multiverse is simulated, the way in which it unfolds in time, and thus
the time variable which will enter the cosmological formalism of \S 2, is determined dynamically. 

To be a bit more precise, the supercomputer can perform operations of three types:
\begin{enumerate}
\item It can advance the surface $\Sigma$ by simulating the laws of physics, leading to a
state defined on a future surface $\Sigma'$, to some approximation obtained by the probabilistic
evolution we just discussed.  We will comment on the cost of this operation shortly.
\item It can make observations at any point to the future of $\Sigma(0)$ and the past of
its current $\Sigma$.  For simplicity we will grant the $4+6$ split, that an observation
is made at a point in $4$-dimensional space-time and returns the type of vacuum $i$,
and that it has a fixed cost.  (We will leave it to future work to assess the plausibility of this assumption.\footnote{Note that we need to simulate a region of the vacuum of sufficient size to
determine $i$ and whether it is hospitable, and the cost of simulating this region may depend on $i$, 
however we count this simulation cost under (1).})
We furthermore require that the observations suffice to deduce whether a vacuum is hospitable
in some fixed amount of time (this is a constraint on the definition of hospitable).
Granting this, as soon as the supercomputer observes that a hospitable vacuum has been
created, it stops and outputs this as the result.
\item Finally, the supercomputer can carry out arbitrary computations of the sort a universal quantum computer
can, with cost given by one of the standard models of computation.  
Perhaps the simplest model, described in textbooks, is the ``quantum gate'' model.
Here one chooses a finite set of finite-dimensional unitary transformations, and the cost
is the minimal number of these ``gates'' needed to approximate the unitary operation which would
exactly implement the computation.
It has also been shown that universal quantum Hamiltonians exist, which act
on a wavefunction in which the program is combined with the computational state
(see \cite{Cubitt:2017} for a recent discussion).
In this case the cost would be the total time of evolution multiplied by the number
of qubits in the wavefunction.
\end{enumerate}
A fully precise definition requires specifying various details in the above formulation including
how the observations in (2) enter the computations in (3).
In the simplest version of the proposal,
we grant that all information gathered by the supercomputer can be combined without regard for
locality in the multiverse, 
and that computations are accomplished by gate operations, each with unit cost.
But one could change these assumptions, say, to enforce locality, or to allow parallel 
or non-deterministic computation. 
In any case, we will seek conclusions which do not depend on such details.

A key point is that  operation (1) of the simulation always reproduces the standard laws of
physics that govern the multiverse.  The additional computational operations (2) and (3)
do not change these laws, they only determine which parts of the totality of the multiverse
actually are simulated.  Thus, we satisfy the principle we stated at the beginning of the 
subsection.\footnote{This requires much more discussion in the quantum case, as measurements
will always lead to correlations which might have observable consequences.  
But if all universes containing observers are semiclassical, this is likely not a problem.}

We now define the global time $t(p)$ of a point $p$ {\it indirectly}, by proposing a
search algorithm using these three types of operations, by which the supercomputer  seeks
a hospitable universe.  Given a run of the search algorithm, the global time $t(p)$ is the computational
time at which the point $p$ is simulated.

Since the simulation operation (1) is probabilistic, we could allow probabilistic computation
in (3), and our measurements in (2) might even have a probability of failure. This implies that
different runs of the supercomputer will generally lead to different outcomes.  
In any given run, once the supercomputer finds a
hospitable vacuum, it stops, and that is the universe predicted by that run.  The probabilistic
aspects of (1), (2) and (3) thereby define a probability distribution over runs, combining the probabilistic
nature of the quantum tunneling events which generate different vacua, with
other probabilistic aspects of the computation.  The resulting measure factor is then the
probability $P_i$ that hospitable universe $i$ is the prediction, under this distribution.

\subsection{Action time}
\label{ss:act1}

Let us begin by stating a useful reframing of our proposal.  Suppose we have a particular 
space-time history for the multiverse on which we need to define a global time coordinate $t$.
Instead of making the details of the search procedure precise, we simply grant that when there
are choices to be made (where to advance $\Sigma$, where to make observations and so on),
these are done in the most efficient way possible.

In this case, 
the global time $t(p)$ of a point $p$, is the {\it minimal} number of quantum operations that a 
quantum computer starting from the initial condition on the hypersurface $\Sigma_0$ would
need to perform in order to simulate the evolution up to $p$.  For this purpose, the computer clearly
needs to simulate every point in the past light-cone of $p$ and to the future of $\Sigma_0$.
Let us call this region $R_p$, and let us denote the number of quantum gate operations required
to simulate the physics in this region as
\be
\CCC(R_p) \equiv \mbox{number of operations to simulate } R_p~. \label{compcost}
\ee

If we fully  understood the fundamental theory, we would know how to compute $\CCC(R)$ for
any space-time region $R$.  We do not, but we can proceed by invoking natural assumptions. 
We expect that $\CCC(R)$
depends only on the physics in $R$, that it can be approximated by the integral of
a local quantity over $R$ perhaps augmented by the integral of another local quantity
over the boundary. 

One simple ansatz is that 
the gate complexity to simulate a space-time region $R$ should be proportional to
its volume in Planck units.  This can be criticized on various grounds, for example it is not holographic.
Recently, a better conjecture has been made by Brown {\it et al} \cite{Brown:2015bva} --
the gate complexity of producing a state on the boundary of $R$
is proportional to the Einstein-Hilbert (or supergravity, or string) action integrated over $R$.
This motivates the following definition:
\begin{definition}
The {\bf action time} $t_a$ of a point $p$ in a cosmology with initial conditions on the hypersurface
$\Sigma_0$, is the integral of the action (in units of $\pi \hbar$) over the region $R_p$.
\end{definition}
We will study this definition in \S \ref{sec:ActionTime}, and argue that it defines a valid time coordinate
in the de Sitter regions of a
cosmology.  This is not immediately obvious, as even when $p$ is in a de Sitter region,
parts of $R_p$ can be Minkowski or anti-de Sitter, but we will establish that this indeed the case.

Action time is the natural definition of global time for quantum cosmology, if we
ignore all of the other costs besides the actual simulation of the physics of the multiverse.  
Thus it is a lower bound for the global time of \S \ref{ss:globaltime} -- but what is its
relation to the search prescription given there?  We briefly discuss this question in \S \ref{s:compclass}; 
it is the search time as defined in a {\bf nondeterministic} model of computation, or as
defined by a ``Merlin-Arthur protocol.''  However we will not use this idea in the rest of this paper,
instead taking the search algorithm to be deterministic.  In this case the discussion is simpler
in a ``local'' perspective, and we now turn to this.

\subsection{Markov formulation of the search algorithm}\label{sec:newmarkov}

We next consider a search algorithm of the general type we discussed in \S \ref{ss:globaltime}, but 
with the further strong constraint that the computational state is entirely composed of the current
state of the simulated multiverse (say, that part of the quantum state which describes the physics
on the hypersurface $\Sigma$), a list of points $p_a(t)$ within the part of space-time which has already been simulated
(including $\Sigma$), a future directed timelike vector at each point $p_a$,
and no additional state.  In this case we argue that the search process
can be described by a Markov process of the same type \eq{markov} derived in previous work on cosmological measure
factors.  Indeed, with some further assumptions, we can derive the new Markov process as a simple
modification of the standard one.

We start by reviewing the derivation of the rate equation \eq{markov} made in 
\cite{Garriga:2012bc}.  The starting point is to postulate a large ensemble of eternal observers (or ``watchers'')
which each follow a geodesic.  We choose a starting point $t=0$ along each geodesic, and define the state
at time $t$ to be the state as observed by the collection of observers each at its own proper time $t$. 
As shown in \cite{Garriga:1997ef,Garriga:2012bc}, the fraction $f_i(t)$ of observers who live in the vacuum $i$
at time $t$ then satisfies the rate equation \eq{markov}, where the rates are the tunneling rates per unit time
per unit spatial volume as described earlier.  Intuitively this is just the statement that when a tunneling between
vacua occurs, an observer who crosses the bubble wall sees the transition, and that these crossings are uniformly
distributed in space-time
(it does not matter that they fall along the same geodesic).

Let us now return to our proposal.  At a computational time $t$, the search procedure must decide where
to advance the current hypersurface $\Sigma$.  By assumption, its only additional state is a list of points
$p_a$, with a new index $a$.  After making observations at each $p_a$,
its options are to choose a point on $\Sigma$ and to advance $\Sigma$ there, or to change the location
of one or more of the $p_a$'s.  Although one can imagine more general possibilities, the simplest one is
to choose one of the points $p_a$ and advance it in the direction of its timelike vector along a geodesic.
If the chosen point is on $\Sigma$, the supercomputer makes the minimal additional simulations required to
enlarge space-time to contain the new value of $p_a$.  The new timelike vector is obtained by parallel transport.

This subset of the possible actions of the search algorithm is both natural, and brings us closer to the
assumptions of \cite{Garriga:2012bc}, as we now have a set of geodesics along which the supercomputer
is making observations.  We also have a natural time variable along each geodesic, but now this is not proper
time but instead the action time evaluated along the geodesic.
Furthermore, rather than follow all of the geodesics and accept the observations along all of them with
equal weighting by point in time and choice of observer, the search algorithm has a variety of other options.
Let us list some of them and explain how they correspond to changing the
Markov process. 

One option is to end the search, if a hospitable vacuum is found.  This makes the hospitable vacuum
terminal and thus it can be represented in the language of the Markov process by setting all the 
transition rates out to zero.

Another option is to duplicate a point on the list.  We give the new point a different timelike vector, so that time evolution will create a distinct geodesic branching off from the first. This is particularly useful if the supercomputer
observes that a transition took place during time evolution, as it can use it to remember the previous
vacuum.  If we always make the choice to duplicate the old point after a transition, 
we can obtain the corresponding Markov process by using the transition matrix
\be \label{eq:defMdup}
M^j_{\ i} = \kappa_{j,i}
\ee
with the second term on the right of \eq{defM} removed.  The duplication could also be done
with some probability $p$, leading to a Markov process in which the second term is multiplied by $1-p$.

Still another option is to abandon a particular geodesic, in other words drop the point $p_a$ from the list.
If we always make this choice for a given vacuum $i$, this amounts to setting all the rates
into $i$ to zero.  Another possibility is to abandon the geodesic with some probability per unit time,
with either a constant rate $r$, or with a rate $r_i$ depending on 
observations which can be made in the vacuum $i$ within some fixed amount 
of computational time.  This can be modeled by adding
a term $-r_i$ to the diagonal entry $M^i_{\ i}$ of the transition matrix.
We will use this idea later for algorithms which
avoid spending time simulating very long-lived vacua.  A similar idea would be to abandon a geodesic
after a fixed amount of time is spent simulating it, but this would be
harder to study as a Markov process.\footnote{We would need to keep the
time spent in each vacuum so far as part of the state.}

The upshot is that many natural search processes can be represented by Markov processes,
with a transition matrix derived in a simple way from the matrix \eq{defM} encoding the transition rates.
We will generally denote this derived transition matrix as $\searchS[M]$, where $\searchS$ is a name for the search
process being considered.  As an example, the simplest search process in this class is the one where we
follow the ``watcher'' prescription, with the only difference that if a watcher sees a hospitable universe
the process stops with this result.  Let us call this process \searchH, then we have
\be \label{eq:Pi-hospitable}
\searchH[M] = M\cdot \left({\bf 1} - \Pi_{hospitable}\right)~,
\ee
where $\Pi_{hospitable}$ is the projection on the hospitable vacua, in other words
a diagonal matrix which acts on the space of number distributions over vacua, is the
identity when restricted to the subspace of hospitable vacua, and is zero on the
complement.  We will discuss the measure factor it leads to in 
\S \ref{s:toymodel}.

As a second example, let us denote as \searchR\ the algorithm which abandons simulating non-terminal vacua with a 
rate $r$, and which treats hospitables as  in \searchH.  Now we have
\be \label{eq:defsearchR}
\searchR[M] = \left(M - r{\bf 1}\right)\cdot ({\bf 1} - \Pi_{hospitable})~.
\ee

A final variation on the theme is the algorithm \searchI.  This is like \searchR\ except that when it
abandons a vacuum, it reverts back to the initial conditions.  The matrix $\searchI[M]$ can be obtained
from $\searchR[M]$ by adding additional matrix elements with rate $r$ for these transitions.

To reiterate, there is an interesting subclass of search algorithms which can be
represented by simple modifications to the transition matrix of the Markov chain.   These are 
algorithms which
observe each new vacuum as it is produced, and decide whether to continue simulating it in
hopes of simulating tunneling events and producing new vacua, or else abandon it and search
elsewhere, based solely on the observed properties of the vacuum itself.  
The simplest of these
properties is simply the lifetime of the vacuum -- if its decay rate is too slow, the supercomputer
may well decide that simulating it is an inefficient way to search compared with simulating other
parts of the multiverse, and act on this decision by switching its attention to these other parts.
These types of algorithms
can be analyzed along the same lines as the existing discussions.  Of course there are many
other algorithms which cannot, because they use the information collected during the search
in more complicated ways; for example there could be an algorithm which ``engineers'' a solution to the cosmological constant problem.  But the simplest algorithms modeled by the Markov processes described above are the easiest to compare with the existing proposals for measure factors.

We can again write the general solution of the rate equation in terms of
eigenvectors and eigenvalues of $\searchS[M]$.  However, 
by contrast, our primary axiom is that cosmological time is limited, so the dynamics may or may not reach equilibrium.  
Thus, the solution will depend on the initial value $N^i(0)$.  Let us write it as
\be \label{eq:exp-growth}
N(t) = e^{\searchS[M]\,t} \, N(0)~.
\ee

Once a hospitable vacuum is found, any given instance of this process will stop.  If we want to
define a probability distribution over the ensemble of such processes, the simplest way to do
this is as follows.  We ensure that the matrix $\searchS[M]$ sets the transition rate out of every hospitable
vacuum to zero -- as we discussed, this represents the ability of the supercomputer simulating
the cosmology to stop and declare this vacuum as the result.  We then take the
limit $t\rightarrow \infty$ of the solution.  With this change, the nature
of the solution can be very different from before, as we will see explicitly in toy
models in \S \ref{s:toymodel}.  We then restrict the final distribution to the hospitable vacua 
(this is only necessary if there are non-hospitable terminal vacua, say supersymmetric Minkowski vacua) and normalize
to get the measure
\be \label{eq:defmeasure2}
P_i = \frac{\lim_{t\rightarrow \infty} N_i(t)}{\sum_{j\ \mbox{ \scriptsize hospitable}} \lim_{t\rightarrow \infty} N_j(t)} ~.
\ee

One might worry that any attempt to define the $t\rightarrow \infty$ limit will lead to the 
difficulties and paradoxes of the standard discussion of measure factors.  On a technical level, one of the
assumptions used to demonstrate these paradoxes is detailed balance, which is spoiled by many of 
the modifications we just discussed.\footnote{Violations of detailed balance associated to AdS bounces
are also part of the arguments made in \cite{Garriga:2012bc}.}
But a more basic argument
that the $t\rightarrow \infty$ limit is well defined in our prescription, is that the volume of space-time
which is being considered is always manifestly bounded because it is related to the 
computational cost of simulating that region.  This is not to say that our proposal is free
of ambiguities, but rather that the ambiguities will be explicit in the definition of the search procedure.  
And since we propose a criterion which favors some search procedures over others -- namely the
search procedure should be as efficient as possible -- there is a new way to try to resolve the ambiguities.

The choice of the transformation $M \rightarrow \searchS[M]$ is a new choice to be made in our approach and to the
extent that it is arbitrary, one might consider this to be a  weakness.  Our answers
to this will be twofold: first, we do not choose the transformation {\it a priori} but rather propose a way
to derive the choice from what we believe to be natural axioms and considerations:
namely, by defining a simulation of the multiverse and a search algorithm which uses the simulation. This will become more clear below.
Second, we will argue that the broad features of the resulting measure factors are not unduly
dependent on the details of this modification.  Moreover, we will argue that natural choices lead to
measure factors which are  very different from 
those which emerge from the standard approach. For instance, in contrast to the standard approach
which favors string compactifications that are 
complicated -- those with few symmetries and much randomness --
the computational limits hypothesis studied here favors such structure as a means of
organizing the search.  
We will also speculate about ways this difference could yield imprints in observable physics.

\subsection{Hospitable vacua and the youngness problem}\label{sec:youngness}
The standard measure program relies on the assumption that we are typical observers. The likelihood of different measurements in our universe (value of $\Lambda$, number of e-folds, time after the big bang, etc.) is calculated from the relative probability of different observers in the multiverse making various measurements for these parameters. If at any point the predictions disagree with our own observations, the measure is ruled out.

For example, a traditional global measure with respect to some time coordinate $t$ counts events prior to a cutoff surface at a constant time $t$. Recall that the probabilities can be computed as follows: Pick a $3$-dimensional region $\Sigma_0$ orthogonal to at least one eternally inflating geodesic (in the end, the answer will not depend on this choice). Construct a series of hypersurfaces $\Sigma_t$ consisting of spacetime points that are located at time coordinate $t$ along a geodesic orthogonal to $\Sigma_0$. Let $N_A(t), N_B(t)$ be the number of observations of type $\mathcal{O}_A, \mathcal{O}_B$ made within the spacetime volume between $\Sigma_0$ and a cutoff surface $\Sigma_t$. Then the relative probability for observers making each of the two measurements is defined as
\be \frac{p(\mathcal{O}_A)}{p(\mathcal{O}_B)} = \lim_{t \rightarrow \infty}\frac{N_A(t)}{N_B(t)}~.\ee
As the cutoff surface is sent to infinity, this approaches a definite value.

It is worth emphasizing that our approach contains an essential difference compared to the standard method. Namely, our simulator searches for \emph{hospitable vacua} that can support life rather than counting individual observers. A hospitable vacuum is defined as a vacuum that the computer has checked to be suitable for life, by simulating some part of its space-time, making observations and using them to test specified criteria, all within a finite specified computational cost.\footnote{Note that a vacuum can be declared``hospitable" even if it does not actually contain life at all.} 
The process of checking for hospitableness figures in as an extra computational cost that contributes to the computation time, Eq.~\eqref{compcost}. The de-centering of observers compared to vacua is one way our measure is fundamentally different from previous measures, and it means our measure is agnostic to certain questions that could be addressed by previous measures, such as the relative probability of being a certain \emph{observer} (say, one who lives a given time after the big bang).

To check for hospitableness, the computer first needs to check that the cosmological constant is small enough to support life. To do so, it must simulate a volume of size $1/\Delta\Lambda$ to measure the c.c. to precision $\Delta \Lambda$. The standard condition~\cite{Weinberg:1987dv} is that the cosmological constant not be so large as to push vacuum domination before the era of galaxy formation: $\Omega_{\Lambda0}/\Omega_{M0} \leq (1+ z_{\rm gal})^3 \sim 100$ for $z_{\rm gal} \sim \mathcal{O}(1)$. Since this estimate requires holding the primordial amplitude of fluctuations $A_{\rm s}$ fixed, a more refined version might involve having the computer also measure $A_{\rm s}$, then having it calculate the effect on the constraint for the cosmological constant.

Since during this check, the computer will detect tunneling events, it is also checking that the vacuum is sufficiently long-lived to produce observers.  This could exclude, for example, groups of nearby minima separated by low barriers which allow frequent intertunneling.  There might be other nonclassical gravitational (and other) phenomena which a good definition of hospitableness would know to check and exclude.  
The definition of these checks might benefit from incorporating  prior knowledge about the string landscape and tunneling rates into the algorithm.  For example, the universal suppression of tunneling events which increase the c.c. and the uniform distribution of c.c.'s imply that tunneling rates out of a low c.c. vacuum are {\bf not} uniformly distributed, so that a vacuum which is stable to
the usual tunnelings available in the string landscape (change of flux or brane configuration) is likely to be very long-lived.

Subsequently, the computer may check any of a number of further anthropic conditions, for example the existence of some sort of chemistry, and a form of energy production like stars. These come with associated computational costs.

The precise choice of what constitutes a hospitable vacuum depends on the algorithm used by the supercomputer. 
Since we insist that this condition can be checked with a fixed computational cost, only a finite amount of space-time
can be simulated.  
This leads to a very different treatment of marginally hospitable vacua, in which observers are possible
but only granting exceedingly rare events.  
For example, in the standard discussions, one needs to consider universes with arbitrarily small average density fluctuations, because there is still a nonzero probability of a sufficiently large density fluctuation to get a region with structure.  The expected number of observers will then be this very small
probability, multiplied by a potentially very large space-time volume in which the density fluctuation is possible.
This is another example of how the factor $X_i$ of \S \ref{s:measures} can influence the measure.

There is an analogous issue with our definition.
Since the results of any given simulation involve a probabilistic (or quantum) uncertainty,
there is a possibility for the hospitableness test to lead to false positives or false negatives.  
For example, a measurement of $A_{\rm s}$ 
will come out with a distribution of values, and even if the mean fails the test, an observed value might pass.
This leads to the analog of the factor $X_i$ of \S \ref{s:measures} --
it is the probability that a vacuum $i$ passes the hospitableness test.
The details depend on the algorithm used to make the test, and are {\it a priori} different from
the conditions for observers to exist.  Because the hospitableness test is supposed to be efficient,
in general the computer makes it at the earliest possible time
(from the point of view of the particular universe under consideration), 
perhaps so early that observers would be exceedingly unlikely to have evolved. 

Without getting into all of the differences this might entail, let us explain how this aspect
of the standard discussion is changed.  In general we still have a chance for a vacuum to be
deemed hospitable due to an exceedingly rare event, quantified by the factor $X_i$.
But for us, the second factor of space-time volume will be cut off in efficient search algorithms, for the same reasons discussed in \S \ref{sec:newmarkov}.  Thus there will be a lower bound on the rate such that 
sufficiently rare events will not affect the final measure.
While the details of this cutoff are algorithm-dependent, this provides a precise and natural way to 
exclude  exceedingly rare events (say with double exponentially small rates) from consideration.

Why the different approach? If the simulation described in \S \ref{sec:supercomputer} was conducted as a search for \emph{observers}, rather than hospitable vacua, it would suffer from an obvious problem: It would predict that the most likely observers are the ones which formed \emph{first} within a single vacuum. This is a kind of youngness problem. Ideally, we would like to limit the computational cost of evolution within the multiverse but not necessarily restrict to the earliest observers within a single vacuum.

  \begin{figure}[t!]
  \centering
  \includegraphics[width=5in]{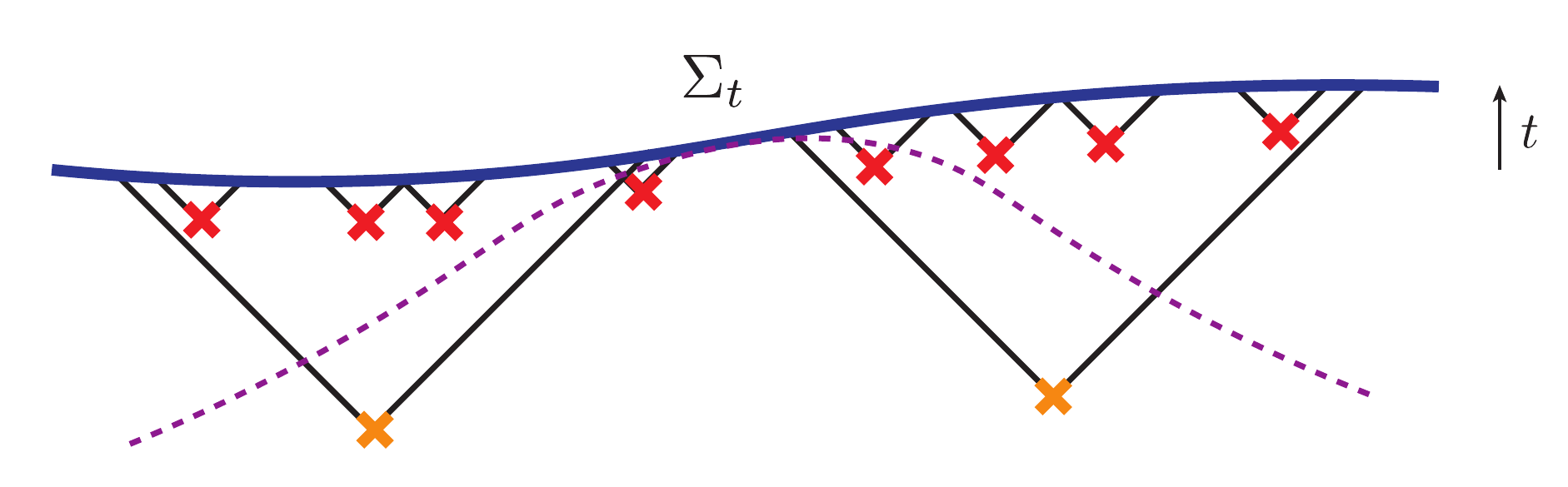}
  \caption{ An illustration of the traditional youngness paradox. For a global cutoff based on a time coordinate $t$ that rewards the exponential expansion at late times, there will be exponentially more vacua nucleated close to the cutoff surface $\Sigma_t$ (marks in red/higher) compared to those nucleated earlier (marks in orange/lower). As a result, most observers will be very young observers that live in the late-nucleated vacua. This contradicts our own observation of the age of the universe. A different choice of time slicing (for example, one with the purple dashed cutoff instead of the solid blue one) would circumvent this problem.
    }
  \label{Youngness}
\end{figure}

Certain traditional measures also suffer from a similar but distinct problem, known as the youngness paradox~\cite{Guth:2007ng, Bousso:2007nd}.\footnote{For an alternate perspective on how to avoid both the Boltzman brain problem and the youngness paradox, see~\cite{Linde:2007nm,Linde:2008xf}.} One example is the proper time measure~\cite{Linde:1993nz, Linde:1993xx}, which is a global measure in the sense we have described above. The essential problem is depicted in figure~\ref{Youngness}. Due to the exponential expansion of space-time, there will be exponentially many more vacua produced closer to a constant proper time cutoff surface. Thus, there will be an exponentially larger number of vacua that were nucleated $< 13.7$ billion years before the cutoff surface than there are older vacua. This exponentially favors very young observers in vacua created at very late times, and is inconsistent with our observations of the age of the universe~\cite{Guth:2007ng, Bousso:2007nd}. As a result, measures like the proper time cutoff are ruled out in favor of  measures based on time slicings which do not reward the exponential expansion at late times. For example, a global measure based on the scale factor time $\eta \sim H \tau$ where $H$ is the Hubble constant circumvents this problem. See figure~\ref{Youngness} for an illustration of how different choices of slicing can avoid the exponential overcounting of young vacua.

Note that the restriction to \emph{hospitable vacua} would not solve the traditional youngness paradox for measures like the proper time cutoff. In fact, we will see that the action time rewards the exponential expansion even more than the proper time (see \S\ref{sec:ActionTime}). Thus, taken as a traditional global measure an action time cutoff would certainly have a youngness paradox of the traditional sense, whether or not we restrict to hospitable vacua instead of observers. By contrast, the computational measure prescription follows local worldlines, so we avoid the exponential proliferation and the traditional youngness paradox as well.

\label{ss:hospitable}

\subsection{Further variations on the search algorithm}

In \S \ref{sec:newmarkov} we considered a variety of simple search
operations, and explained how they can be reformulated in terms of a Markov 
process.  Here we combine these to obtain some simple search algorithms, which
we will go on to study in more detail in later sections.  There are also many
search algorithms which cannot easily be reformulated as Markov processes,
and we will describe a few of these.

Let us begin by discussing the simplified version of \S \ref{ss:act1} in which the search
procedure is abstracted away to the claim that it is done ``as efficiently as possible.''  In this case,
 global time is action time.  As we will show later, this is more or less \eq{defbetatime}
with $\beta=2$.  Intuitively, this is because vacua with large c.c. are ``simpler'' and thus easier
to simulate.  The search is then done by evolving the surface $\Sigma$ forward at random
points (distributed uniformly in the spatial volume measure).  Thus there is no additional structure
besides the standard cosmological dynamics, and the resulting Markov process will be the standard
one with transition matrix \eq{defM} with $\beta=2$.

If we take this as the final result, without any restriction to hospitable vacua,
then of course we should run into the same paradoxes as
the standard discussion.  This is consistent with our claims because this measure is the
result of a nondeterministic search algorithm, so the volume of potentially simulated space-time as 
a function of time can grow far more quickly (nondeterminism adds at least another exponential).

What about a nondeterministic search which stops at the earliest hospitable vacuum?
We believe that this will not suffer from these paradoxes.
On the other hand, the resulting measure factor is much more likely to depend on the initial conditions.
We will argue in \S \ref{ss:dependence} that this will not be a dependence on details, but rather on general properties
such as ``is the initial condition supported on simple vacua''?

Let us turn to versions of the proposal with a deterministic search algorithm.  Thus, rather than blindly
simulate the entire multiverse, the supercomputer can use its partial results to guide its future search.
The simplest case of this is to allow it to detect hospitable vacua and then stop.  As we explained in
\S \ref{sec:newmarkov}, the effect of this is to choose a subset of vacua -- here the hospitable ones -- and
by fiat make them terminal.

Going beyond this to improve the search algorithm, we must grant some ability for the supercomputer
to decide which vacua are promising candidates for time evolution, and which are less promising.  One can imagine many different assumptions about the prior knowledge of the string landscape which is built into the program.  At the one extreme, one could imagine that its structure has been completely understood mathematically and that this knowledge has been used to produce an optimal algorithm to produce hospitable vacua.  This might or might not lead to the ``most efficient'' version of the proposal we already discussed.  At the other extreme, we could give the supercomputer no prior knowledge of specifics, only
generalities at the level we are discussing in this paper.  This is obviously an easier case to discuss, so let us focus on it.

Compared to a random procedure for simulating the multiverse, the most obvious scope for optimization is to not simulate very long-lived vacua.  Rates which come out of the formula \eq{cdlrate}  are often very
small, especially for upward transitions from low c.c. vacua because of the detailed balance formula \eq{upward}.  Of course low c.c. vacua must be explored because they have a chance to be hospitable, but that is not to say that the supercomputer must devote vast resources to their exploration.  An efficient search procedure would probably choose a cutoff time $T_{cutoff}$, longer than the time needed to determine hospitability, and perhaps longer still for other reasons, and only simulate until that amount of computational time is exhausted.  If it is exhausted, the search will continue by evolving the space-like hypersurface from some other vacuum.   We will speak of the supercomputer as ``abandoning'' that vacuum.

We have described a version of this idea already in \S \ref{sec:newmarkov}.
To get a Markov process, rather than a cutoff time $T_{cutoff}$, we postulated a cutoff rate $r$.
For $r\sim 1/T_{cutoff}$ the results should be similar.
A variation is to duplicate the simulation point $p_a$ before each transition, providing some ability to
backtrack.  Of course if we were willing to complicate the Markov process, we could keep track of which
geodesics the points $p_a$ belong to, and then when a vacuum is abandoned go back to its ancestor.  

Another version would be for the cutoff transition not to eliminate the currently simulated universe
(at the point $p_a$) but to instead force a transition to the initial condition (equivalently, move
$p_a$ back to the initial hypersurface).

Another interesting variation on the proposal is an ``exploration'' version in which the goal of the
supercomputer is not to find a single hospitable vacuum, but rather to more efficiently implement
the cosmological dynamics which generates a large set of distinct vacua.\footnote{
This idea arose in discussions with Raphael Bousso and Alan Guth.}  To define this variation,
we do not choose the $P$ matrix to make the hospitable vacua terminal, but we incorporate one of
the modifications we just discussed to optimize the search by abandoning long-lived vacua.
We then restrict the measure to the hospitable vacua before interpreting the result: thus the
final measure factor is \eq{defmeasure}.

Finally, we raise the possibility that there is structure in the string landscape which could be
explicitly used to speed up the search.  As an example, suppose we could find a class of Calabi-Yau
compactifications with many independently adjustable fluxes $N_a$, such that each flux contributed
vacuum energy $N_a^2 M_{flux}^4 E_a$ for a hierarchically arranged set of energy scales 
$E_a = 1,10^{-1},10^{-2},\ldots,10^{-120}$.  This would facilitate a search which systematically
solved the cosmological constant problem on successively smaller scales.  The possibilities for such
``engineered'' corners of the landscape are many, and in the conclusions we discuss some testable
consequences were this to be the way cosmology worked.

\section{Models of the landscape}
\label{s:toymodel}

In this section we study some simple models of the landscape to illustrate our proposal.  
We begin in \S \ref{ss:toymodel1} with a variation of a model studied in \cite{Bousso:2012es}.
It contains four vacua, of which two are hospitable and one is long-lived.  We will see that
the modified Markov processes discussed in \S \ref{sec:newmarkov} behave as expected, in
particular that cutting off long-lived vacua drastically reduces the mixing time.

In \S \ref{ss:toymodel2} we study a slightly more complicated toy model, with additional
vacua labeled by a ``complexity'' axis.  We will use this to
study the dependence on initial conditions.

\subsection{Toy Model 1 -- effects of modifying the Markov process}
\label{ss:toymodel1}

\begin{figure}[t!]
  \centering
    \includegraphics[height=0.4\textwidth]{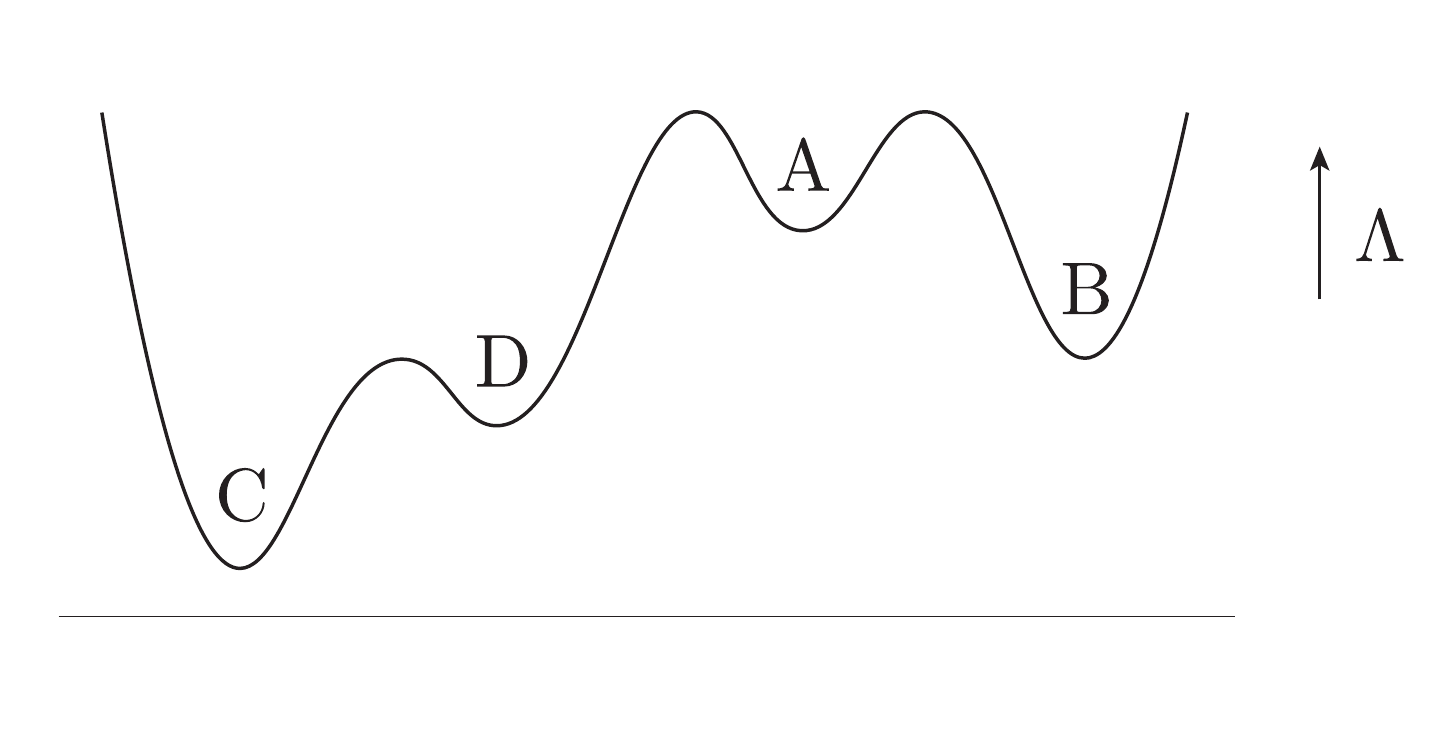}
    \caption{The potential landscape for toy model 1.} \label{fig-toymodel}
\end{figure}

The first toy model is depicted in figure~\ref{fig-toymodel}. 
The two hospitable vacua are $D$ and $B$, the initial conditions start in $A$, and $C$ is long lived (the ``dominant" vacuum in the standard approach).  We may also have decays to other terminals (say Minkowski or AdS) not shown in the figure. For simplicity, we have only allowed for decays between neighboring vacua and from $C$ to $A$.

The vacua are ordered by increasing cosmological constant, $\Lambda_C < \Lambda_D < \Lambda_B < \Lambda_A$. 
The decay rate from vacuum $i$ to $j$ is denoted $\kappa_{ji}$, and total decay rates are $\kappa_i$.  In the absence of terminals the total decay rates would equal
the sum of individual decay rates out of the vacuum $i$ to all the other de Sitter vacua pictured.
They will in general be larger than that sum if there are additional decays to terminals.

Let $M$ be the transition matrix in the standard eternal inflation rate equation \eq{markov},
and $\searchH[M] = M \cdot ({\bf 1} - \Pi_{hospitable})$ be the transition matrix modified 
as in \eq{Pi-hospitable} to make hospitable vacua terminal.  We take
\be
M = \left(\begin{array}{cccc}
-\kappa_C & \kappa_{CD} & 0 &0\\
\kappa_{DC} & -\kappa_D & 0 & \kappa_{DA}\\
0&0& -\kappa_B & \kappa_{BA}\\
\kappa_{AC} & \kappa_{AD} & \kappa_{AB} & -\kappa_A
\end{array}\right)~,\indent
{\bf 1} - \Pi_{hospitable} = \left(\begin{array}{cccc}
1 & 0 & 0 & 0\\
0 & 0 & 0 & 0\\
0 & 0 & 0 & 0\\
0 & 0 & 0 & 1\\
\end{array}\right)~.
\ee
By assumption $C$ is long lived, so all of the rates $\kappa_{iC}$ and $\kappa_{C}$
will be very small, typically double exponentials.

Let us first analyze the equilibrium solution of the standard equation along the lines of SPV.  
They observed that, because of the
detailed balance factor \eq{upward}, in realistic landscapes with small c.c. vacua,
upward tunneling rates are generally much smaller than the corresponding
downward rates.  This suggests splitting the transition matrix into an upper triangular downtunneling part
and a lower triangular uptunneling part,
\be \label{SVPsplit}
M = M_{down} + M_{up}~,
\ee
and perturbing in the uptunneling part.  Note that we partition the diagonal entries into downtunneling
and uptunneling parts as well,
\be
-\kappa_i = - D_i - U_i~.
\ee

In the toy model, the dominant eigenvalue is approximately equal to $-D_C$, and the 
dominant eigenvector will have a large $C$ component, which we can take to equal $1$.
The other entries of the dominant eigenvector can be computed perturbatively: the entry
corresponding to vacuum $i$ is a sum of products of the ratios $\kappa_{ji}/(D_j - D_C)$ 
along all paths from the dominant vacuum $C$ to the vacuum $i$.  

If we replace $C$ of the toy model with the longest-lived non-terminal vacuum,
essentially the same results govern the equilibrium limit of very general rate matrices coming out of
eternal inflation.  This is summarized in the phrase ``the measure is given by tunneling rates from the
longest-lived metastable vacuum.''

The mixing time is set by the gap between the first and second eigenvalues, and the second eigenvalue is set by the second smallest decay rate plus corrections. While the second eigenvalue will be large in the toy model,
in a more realistic large landscape many decay rates will be double exponentially small, so it will be typical for the mixing time to be very large, of order the second smallest decay rate.

Let us now turn to the computational models, beginning with $\searchH$.  Besides the modification to
the transition matrix, we are now instructed to compute the measure factor \eq{defmeasure2}
from the long time limit of the restriction of $f_i(t)$ to the hospitable vacua.  This is not directly related
to the dominant eigenvector, but we can find it from an eigensystem analysis.  To do this we write
the general solution as
\be \label{eq:foft}
{\bf f}(t) = {\bf f}_\infty + \sum_i C_i {\bf s}_i e^{-q_it} 
\ee
where $({\bf s}_i,q_i)$ are the eigenvector, eigenvalue pairs and ${\bf f}_\infty$ is supported only on the terminal vacua.
We then determine ${\bf f}_\infty$ and the coefficients $C_i$ by matching the initial conditions ${\bf f}(0)$.

As one might expect, this prescription has far more scope for dependence on the initial conditions, and
we will explore this dependence below in models with more vacua.
On the other hand, it can be far less dependent on the ``inhospitable'' parts of the landscape and the
long-lived vacua.  In the model at hand, one can easily show that ${\bf f}_\infty$ has entries $c_D, c_B$ with
\be 
c_D = \frac{\kappa_{DA}}{\kappa_A}~, \indent c_B = \frac{\kappa_{BA}}{\kappa_A}~.\ee
In other words, the relative weight in the measure is the set by the branching ratios out of $A$. 

In general the mixing time is still set by the gap or equivalently the smallest absolute magnitude of the $q_i$.
In search algorithm \searchH, 
the smallest absolute magnitude eigenvalue will again be approximately $-\kappa_C$, so in this toy model the mixing time is still long.
However, there are more efficient search algorithms than \searchH, which avoid simulating
very long-lived vacua.  In search algorithm \searchR, we modify the transition matrix to \eq{defsearchR},
so now $\kappa_C$ is no longer small, and the mixing time becomes of order $1/r$.
While trivial mathematically, this makes the point at hand.

\subsection{Toy Model 2 -- dependence on initial conditions}
\label{ss:toymodel2}

The next point we want to study is dependence on initial conditions.  On general grounds, and as we
saw in toy model 1, the computational measures will almost inevitably depend on initial conditions.
Now from our present standpoint where we know nothing about the initial conditions,
this is a deficiency.  On the other hand, 
if we actually knew something about initial conditions
which naturally come out of string theory, this dependence could give us a fairly direct path towards making predictions. As we discuss in \S \ref{ss:dependence}, we posit that these initial conditions are ``simple."

Let us suppose this is the case.  To be a bit more concrete, let us
label compactifications by a topological number $b$, the ``number of cycles'' or ``number of fluxes.''
This could be the sum of the Betti numbers in Calabi-Yau or $G_2$ compactification, and might also take into
account other sectors such as branes which have parameters which vary between vacua.  
In the Bousso-Polchinski model, $b$ will be the dimension of the flux lattice.
We suppose that $b$ takes values from $b=0$ up to some finite $b=b_{max}$.  

Now, if the initial conditions were concentrated on $b=0$, would
this give us any reason to think that the measure factor for hospitable vacua will be concentrated on small $b$?
In a prescription without dependence on the initial conditions, clearly the answer is no.
What about a prescription like the computational measure, or another non-equilibrium measure?

One fact about the landscape which is very relevant for this discussion is that there are far more vacua
with large $b$ than small $b$.  This is intuitively plausible on combinatorial grounds, and it
is postulated in the Bousso-Polchinski model, where each flux can
take any of a bounded quantized set of values $|N^i|<N^i_{max}$ before the cosmological constant becomes too large.
We can model this dependence by the formula $N_{vac}(b) = (2N_{max})^b$.
It can be derived (in the regimes where it holds) from more accurate vacuum counting formulas 
such as those in \cite{Ashok:2003gk,Denef:2004ze}.

Because of this fact, entropic considerations favor complicated vacua.  This might be the case even for
a prescription which depends on initial conditions, so the answer to our question is by no means obvious.
Let us extend our toy model so that we can study it.

\begin{figure}
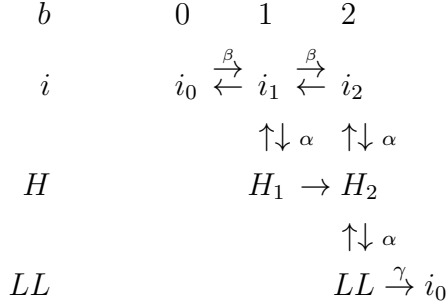
 
\begin{align}
b \qquad \qquad &0 \qquad \,1 \qquad \,2 \nonumber\\
i \qquad \qquad &i_0 \,\overset{\overset{\beta}{\mbox{$\rightarrow$}}}{\leftarrow}\, i_1 \,\overset{\overset{\beta}{\mbox{$\rightarrow$}}}{\leftarrow}\, i_2 \nonumber\\
 &\hspace{1cm} \uparrow\downarrow \mbox{\scriptsize$\alpha$} \hspace{0.25cm} \uparrow\downarrow \mbox{\scriptsize$\alpha$} \nonumber \\
H \qquad \qquad & \qquad \ H_1 \,\rightarrow H_2 \nonumber\\
&\hspace{1cm} \hspace{1.1cm} \uparrow\downarrow \mbox{\scriptsize$\alpha$} \nonumber \\
LL \qquad \qquad & \hspace{2.1cm} LL \overset{\gamma}{\rightarrow} i_0\nonumber  
\end{align}
\caption{The vacuum decay chain of toy model 2.} \label{fig:modeltwo}
\end{figure}

Our second toy model is a decay chain depicted in figure \ref{fig:modeltwo}.
We have taken $b_{max}=2$, and we will postulate that at
each fixed $b$ there are $N^b$ vacua for some fixed $N$. 
There are two hospitable vacua, $H_1$ and $H_2$ and a single long lived vacuum $LL$. 
The initial condition is set at $i_0$.

The transitions $b\rightarrow b+1$ have a rate set by the parameter $\beta$, whereas the inverse decay $b \rightarrow b-1$ has rate $\beta/N$. Downward tunnelings in $\Lambda$ are taken to have fixed rate $\alpha$ and uptunnelings are further suppressed by a factor $f$ which we take constant for simplicity.   We have also allowed for a fixed large decay rate $\gamma$ which would implement the search algorithm \searchI \ by returning to the initial conditions $i_0$ from $LL$. 

In an equilibrium framework, the measure is controlled by the long lived vacuum $LL$.
And as we discuss at more length below, one expects $LL$ to have large complexity,
both on entropic grounds and because there is more scope to adjust the parameters to favor a long lifetime.
In the toy model, we simply assign it to $b=b_{max}$.  Then the considerations discussed in the previous subsection
show that in the equilibrium measure,
\be \label{eq:equil-measure-example}
\frac{P_1}{P_2}\bigg|_{equilibrium} = \frac{\beta}{\alpha N}~.
\ee
The complexity of the dominant vacuum favors complex hospitable vacua.

Let us turn to the computational measure with search algorithm \searchI.
The relevant matrices are
\be
M = \left(\begin{array}{cccccc}
-\gamma & \alpha & 0 & 0 & 0 & 0 \\
0 & -(1+f)\alpha & 0 & \alpha & 0 & 0\\
0 & 0 & -f\alpha & 0 & \alpha & 0\\
0 & f \alpha & 0 & -(\alpha + \frac{\beta}{N}) & \beta & 0\\
0 & 0 & f\alpha & \frac{\beta}{N} & -(\alpha + \beta + \frac{\beta}{N}) & \beta\\
\gamma & 0 & 0 & 0 & \frac{\beta}{N} & -\beta
\end{array}\right)
\ee
with the projection ${\bf1}-\Pi$ a diagonal matrix of $1$'s except for zeros in the second and third entries corresponding to the vacua $H_2, H_1$ respectively. 

The qualitative behavior of this model is controlled by the ratio
\be
c \equiv \frac{\beta}{\alpha} 
\ee
which is a suppression factor for topology changing transitions compared to other vacuum tunnelings.
We may speculate that transitions that change $b$ are rare, occurring for example only near special conifold points, so that $c \ll 1$.  
Additionally, this particular model is easiest to solve expanding around $c=0$ so we will restrict to that case.

The full set of eigenvalues up to first order in $c$ are:
\begin{align}
q_0&= 0~,\\
q_1&=0~,\\
q_2&=-\gamma~,\\
q_3&=-\left[1+\left(\frac{1}{2}+\frac{1}{N}-\sqrt{\frac{1}{4}+\frac{1}{N}}\right)c\right] \alpha~,\\
q_4&=-\left[1+\left(\frac{1}{2}+\frac{1}{N}+\sqrt{\frac{1}{4}+\frac{1}{N}}\right)c\right] \alpha~,\\
q &= -c \,\alpha~.
\end{align}
The smaller of $\gamma$ and $\beta$ will be the dominant eigenvalue which controls the mixing time.

To find the weight in the measure, we can again apply \eq{foft} evaluated at the initial conditions. 
This is a long calculation but it can be done in an expansion in $c$.
We find that to leading order, 
\be  \label{eq:comp-measure-example}
\frac{P_1}{P_2}\bigg|_{\mbox{search }\searchI} = \frac{1}{c}~.
\ee
Thus this measure favors $H_1$ (the less complex hospitable vacuum) over $H_2$ by a large factor. 

Let us compare this with the same model but with ``complicated'' initial conditions $i_2$. 
In this case the leading order contribution is
\be  \label{eq:comp-measure-example2}
\frac{P_1}{P_2}\bigg|_{\mbox{search }\searchI} = \frac{c}{N}~.
\ee
So the measure has changed back to agree with the equilibrium result
\eq{equil-measure-example}.

We can understand these results intuitively by looking at figure \ref{fig:modeltwo} and comparing the shortest
paths from the initial condition to each of the hospitable vacua.  The idea is to look at the points at
which the paths diverge and find the relative fraction of the total rates which take each path.  Starting at $i_0$ the paths branch at $i_1$, with relative probabilities $(\beta,\alpha)$ to make
transitions to $(i_2,H_1)$ respectively.  
This leads to \eq{comp-measure-example} (at leading order in $c$).

This picture can be generalized to more complicated landscapes as follows. Namely, for an initial condition concentrated on
vacuum $i$, we expect that\footnote{
We believe this formula may be exact if there are no other terminals, but 
we are still studying this question.
} the large time limit of the measure for hospitable vacuum $j$ 
is the sum of a term for
each path from $i$ to $j$ to be 
given by a product of branching ratios along that path,
\be \label{eq:path-sum}
(f_\infty)^j =  \sum_{\mbox{paths } i\rightarrow j} \prod_x \frac{\kappa_{y,x}}{\sum_z \kappa_{z,x}}~.
\ee
Here the product is over intermediate vacua $x$ such that there is more than one outward transition which could lead to a hospitable vacuum.

According to this picture, the long lived vacua play no role.  This will indeed be the case
in search algorithms such as \searchR\  in which 
the transition matrix has a diagonal term $-\kappa_{LL}=-r$ much larger than
the other decay rates of $LL$, as then the $LL$ component
of the probability is negligible.  In algorithm \searchH, we must consider paths in \eq{path-sum}
which go through $LL$, but by assumption these will involve a very small weight and can be
neglected for this reason.  

On the other hand, if long-lived vacua are taken to recycle into special vacua as in
algorithm \searchI, they could have a significant effect on the measure factor.  But since in 
\searchI\ we take these special vacua to be the same as the initial vacua, we do not expect
this to change the measure factor either.

The upshot of this analysis is that the computational measure is in general very different from
an equilibrium measure -- not just in the results, but in having significant dependence on the
initial conditions, and in that the Markov process takes much less time to reach the final results.
This was true for a variety of search algorithms, and the variations between algorithms that we
considered did not lead to significant changes in the final results.

\section{Application to the string landscape}
\label{s:measureresults}

In \S \ref{s:toymodel} we looked at toy models of the landscape and showed that 
the modifications to the Markov process which come out of the computational prescriptions change
the measure factors.  We pointed out the following differences with the equilibrium measure:
\begin{itemize}
\item Both the equilibrium measure, and an interesting subset of the computational measures,
can be formulated as the large time limits of Markov processes.
For both, the time needed to approach this limit is the inverse spectral gap.
Whereas in the equilibrium measure this is controlled by the longest lifetimes of metastable vacua,
in the computational measure it is controlled by parameters of the search algorithm.
\item The equilibrium measure is controlled by the dominant eigenvector and thus favors vacua
which can be easily reached from the longest lived metastable vacuum.  By contrast, the computational
measures are generally controlled by the initial conditions and favor vacua which are close to the
initial conditions.  The long-lived vacua will generally play no role, both because their lifetimes will be 
cutoff in efficient search algorithms, and because there is insufficient cosmological time for
the chains of tunnelings which lead back from long-lived vacua to hospitable vacua to contribute.
\end{itemize}
We believe that these claims hold
for more general and complicated landscapes.  Rather than try to exhibit them in more detailed
and complicated toy models, we will try to extend them to the actual string landscape.

\subsection{Role of long lived vacua}

As we argued, the longest-lived vacua which play an important
role in the equilibrium prescription need play no role here.  Quite the contrary. In the string landscape one expects
the longest-lived vacua to come from compactification manifolds with complicated topology and
a very large number of homology cycles~\cite{Douglas:2012bu,Taylor:2015xtz}.
Thus they would require a long time to be created through cosmological dynamics from any simple initial conditions.
And so it is reasonable to expect hospitable vacua to be generated well before these long-lived vacua would be, in which
case they would largely be irrelevant.

More explicitly,  suppose that in the string landscape there is a fairly short chain of tunneling events which, starting from some simple initial vacuum state, can generate
the required structure in the extra dimensions to realize the Standard Model, the small cosmological constant and whatever
else a vacuum needs to be hospitable.  In our prescription, this would clearly be a favored candidate
for our vacuum.  But in the equilibrium prescription, we are instructed to ignore the fact that this vacuum
is easy to create; rather the multiverse is dominated by the longest-lived metastable de Sitter 
vacuum. The simple vacuum configuration will only appear if it can be easily reached from that vacuum.  And so, the standard picture
favors complicated vacua that can be easily reached from the dominant vacuum.

Although logically consistent, the equilibrium picture is counterintuitive and almost certainly would 
not be considered the natural prediction of string theory.  Occam's razor and many other considerations
would tell us to favor the vacua which are simple in their internal structure and simple
to create. Moreover, the most emphatic criticism of the usual multiverse picture is that it postulates an
almost unimaginable plethora of unobservable universes.  While the theory of string compactification
combined with the dynamics of early cosmology does predict the existence of unobservable universes, one can surely ask for a paradigm which postulates a minimal number of such  unobservable universes.  Our new prescription explicitly addresses this point and is a minimality condition.  However, in our prescription, the quantity being minimized is
not the number of universes, but the resources it would take to create or simulate them.

\subsection{Simple initial conditions}
\label{ss:dependence}

One might worry that the significant dependence of the computational prescriptions on initial conditions
will preclude ever making any predictions from them.  After all, we do not know the initial conditions.

We now make two arguments to counter this.  First, while we do not know whether string theory favors
particular initial conditions, there are reasons to think that some string compactifications are simpler
than the others and that these will be the favored ones.  Second, while the statement we just made does
not sound very precise, we will argue that we do not need a very precise result for the initial conditions;
we only need to narrow them down within a subset of vacua which are connected by relatively fast
transitions compared to those which are required to produce hospitable vacua.  

At least from the point of view of a practicing string theorist, the claim that some string compactifications are simpler
than others seems manifest.
There are good reasons why conjectures in string compactification are first studied with torus and orbifold
target spaces, and then with complete intersection Calabi-Yau manifolds or other relatively concrete manifolds.
These are produced by simple mathematical operations from manifolds with large symmetry groups and
are thus easier to analyze, and will remain so no matter how sophisticated our understanding may become.

Another measure of the complexity of a compactification manifold is in terms of the Betti numbers (as in our
toy models), or in terms of the number of equations we need to describe it.  We can then count the number
of additional parameters -- fluxes, brane configurations and the like -- needed to uniquely specify the
vacuum.  If we need to give the initial conditions to the supercomputer, surely this
will be simpler for a smaller number of parameters. 

Both of these measures can be combined by saying that we can measure the complexity of a compactification 
by the minimal length of the description needed to specify it in some formal language.  
This measure is known as Kolmogorov complexity.  
Of course we are making an additional axiom to say that the initial conditions should have small Kolmogorov complexity,
however this is an axiom which fits well with the other axioms in our computational approach.

A superficially diffferent approach to the question would be to say that solutions are simpler
if they have more symmetry.  A related idea is that mathematical classifications often reveal the
existence of ``exceptional'' objects, such as $E_8$ in the theory of Lie algebras,
and one can easily imagine that some string solutions will be exceptional in an
analogous sense.  We suggest that
this approach is only superficially diffferent as one should then be able to characterize the object using
its symmetry or its other exceptional properties.  Of course, string theory might
prefer exceptional solutions for the initial condition for its own internal reasons,
just as it prefers $E_8\times E_8$ and $Spin(32)/\IZ_2$ as ten dimensional gauge
groups.

Indeed, if we had a quantum nonperturbative definition of string theory, it seems likely that particular solutions
would appear as simpler than others, as is the case in our other physical frameworks.  We could draw an
analogy to the discussion of initial conditions in observable cosmology.  It is not obvious what
configurations of electrons, protons, photons and so forth we should call ``simple,'' except for thermal
configurations which have reached equilibrium.  But this is not really the right question; rather the initial
conditions for cosmology come out of the vacuum configuration for a quantum scalar field, through the
dynamics of inflation and reheating.  Perhaps the question of which compactification geometries are
preferred, will be answered by finding some more basic level from which geometry emerges.

\subsection{Independence of details of initial conditions}

The essential point here is already visible in our toy models.  It is that a group of vacua connected by
relatively fast tunneling rates will equilibrate on a time scale set by these rates.  It is reasonable to
expect that simple vacua will have large vacuum energy and relatively fast tunnelings, compared to
the low c.c. hospitable vacua and the tunnelings required to produce more complicated vacua.

There are reasons to believe this will \emph{not} be true in a simplified Bousso-Polchinski landscape~\cite{Bousso:2000xa}. This model consists of a $J$-dimensional lattice of flux vacua labeled by a vector of integers $N\in \mathbb{Z}^J$. 
The vacuum energy of a given vacuum in this landscape is

\be \Lambda = \Lambda_0 + \sum_{ij} g_{ij} N^i N^j~,\ee
where $\Lambda_0$ is the bare cosmological constant and $g_{ij}$ is a positive definite metric.

Starting from a ``simple" initial condition such as a compactification along a $T^6$ corresponds to picking an initial vacuum with a large cosmological constant set by the Planck scale, $\Lambda_{\rm init} \lesssim M_P^2$. Such initial conditions are spread out along a large shell with radius $\Lambda_{\rm init}$, while the anthropic vacua lie along a shell of fixed $\Lambda_{\rm us} \ll \Lambda_{\rm init}$ with width $\Delta \Lambda_{\rm us}$. In general down-tunnelings are relatively unsuppressed compared to up-tunnelings. Given exponentially many anthropic vacua, there will be an exponentially large number of different unsuppressed paths that (in general) begin at different starting vacua with $\Lambda_{\rm init}$. We do not have a reason to expect that any one of these paths be dominant over the others, and thus the calculation seems to depend considerably on which ``simple" initial vacuum is chosen. Indeed, we have checked this behavior numerically for $\mathcal{O}(500)$ vacua. As the dimensionality of the landscape is increased, the number of hospitable vacua and thus paths will only increase; additionally, there is evidence that down-tunneling from a high to low c.c. becomes even faster as the dimensionality of the landscape is increased~\cite{Greene:2013ida}. Thus, we expect the dependence on initial conditions to become even worse for realistically large landscapes.

However, we believe that this feature is a consequence of the simplifications inherent in the Bousso-Polchinski model rather than a property of the actual string theory landscape. In fact, we can consider a slightly more complicated model which allows for flux-changing transitions, much along the lines of the second toy model considered in \S\ref{ss:toymodel2}. We can imagine that complexity (the parameter $b$ in the toy model) is set by the total charge of the solution. For example, in type IIB compactifications this would be the spacetime-filling D3-charge stored in fluxes or in D3-branes, which is set by the Euler characteristic of the Calabi-Yau 4-fold (which produces a background $Q_3 = - \chi/24$). Transitions that increase this charge would require a tunneling event creating at least one D3-anti-D3 pair filling a sufficiently large patch of space, followed by the conversion of the anti-D3 to background curvature (increase of Calabi-Yau 4-fold Euler characteristic by 24 units) through a topology-changing transition. It is plausible that such transitions are highly suppressed.

By this reasoning, we suggest that string theory may exhibit a ``highway" of simple vacua connected via unsuppressed tunnelings, with paths out of this highway highly suppressed compared to the rates to convert between simple vacua. Furthermore, given the exponential and double exponential nature of typical decay rates in the landscape, we imagine that even given exponentially many paths to anthropic vacua there may be one which clearly dominates. Such a path begins in some vacuum $\Lambda_i$ within the highway. If we choose a different initial vacuum $j$ inside the highway, the path that consists of tunneling (unsuppressed) from $j$ to $i$, and then tunneling along the dominant path from $i$ out of the highway to an anthropic vacuum will be much faster compared to directly tunneling out of the highway from vacuum $j$. Thus, the first anthropic vacuum along this dominant path is selected, regardless of whether the initial conditions are chosen to be vacuum $i$ or vacuum $j$. Assuming such a highway and a dominant path out, the anthropic vacuum selected by the approach should not depend on the starting point.

This can be contrasted with the standard approach, where the initial starting point is instead a small $\Lambda$ ``dominant" vacuum with the slowest decay rate. The decay path to an anthropic vacuum will start with a suppressed up-tunneling out of this dominant vacuum, followed by further tunnelings (in general down-tunnelings) to reach a viable vacuum. Depending on which vacuum one starts with, this uptunneling may land in different anthropic regions and give a different result.

\section{Results on action time}\label{sec:ActionTime}

A simple choice of global time coordinate that serves as a clock for our simulator was presented in \S \ref{ss:act1}. Here we would like to elaborate on this choice. This section can be safely skipped by non-experts. 

\begin{figure}[t!]
  \centering
  \includegraphics[width=5.5in]{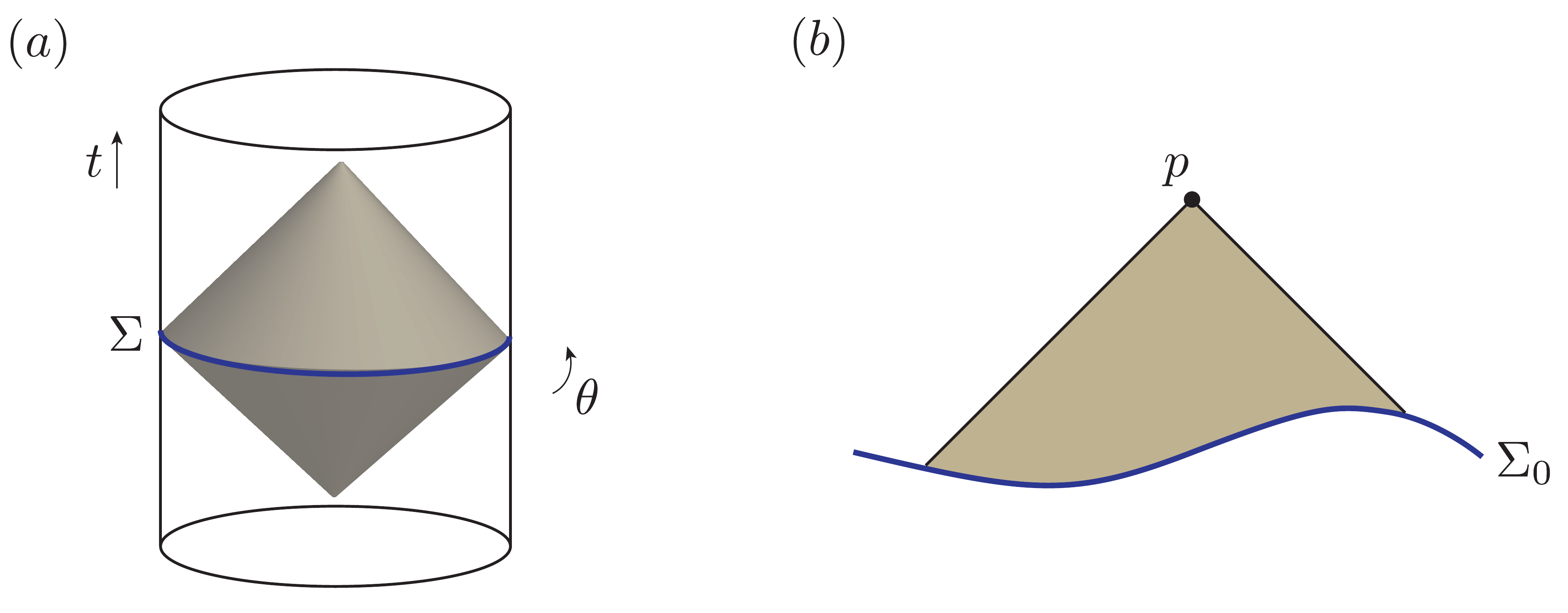}
  \caption{(a) The Wheeler de Witt patch corresponding to a $t=0$ boundary Cauchy slice $\Sigma$ in pure AdS. The action integrated over this region is conjectured to be dual to the complexity of the boundary state on $\Sigma$.  (b) In the multiverse we introduce an ``action time" coordinate $t(p, \Sigma_0)$, defined as the integral of the gravitational action over the causal past of a bulk point $p$ intersected with the future region of influence of a bulk Cauchy slice $\Sigma_0$. We conjecture that the elapsed action time computes the computational cost of simulating bulk evolution from $\Sigma_0$ to a surface of constant $t(p, \Sigma_0)$. 
    }
  \label{ActionTime}
\end{figure}

The complexity of a state is (roughly) the number of universal quantum gates needed to prepare the state from a simple reference state using a quantum circuit. For theories which admit a semiclassical asymptotically anti de Sitter (AdS) dual, there have been two recent holographic proposals for a gravitation interpretation of complexity. First, the complexity = volume conjecture suggested that boundary complexity may be computed by the volume of an appropriate bulk time slice~\cite{Susskind:2014rva}, up to an arbitrary length scale. Most recently, the complexity = action conjecture~\cite{Brown:2015bva, Brown:2015lvg} relates the complexity $\mathcal C$ of a state on a time slice $\Sigma$ to a bulk gravitational action:

\be \mathcal C(\Sigma) = \frac{I_{\rm WdW}}{\pi \hbar}~.\ee 
Here $I_{\rm WdW}$ is the gravitation action including a cosmological constant integrated over the Wheeler de Witt patch, which is defined as the domain of dependence of any bulk Cauchy slice that asymptotically approaches $\Sigma$ on the boundary (see figure~\ref{ActionTime}a). 

In an eternally inflating cosmology, a putative dual theory is believed to live on a time-slice at future infinity. Many of the details of this ``correspondence," including exactly what the dual theory is, are poorly understood. Nonetheless, it seems reasonable to posit an analogous relation between bulk action and complexity, this time of \emph{bulk} states that are related by cosmological evolution. 

\begin{figure}[t!]
  \centering
  \includegraphics[width=4.5in]{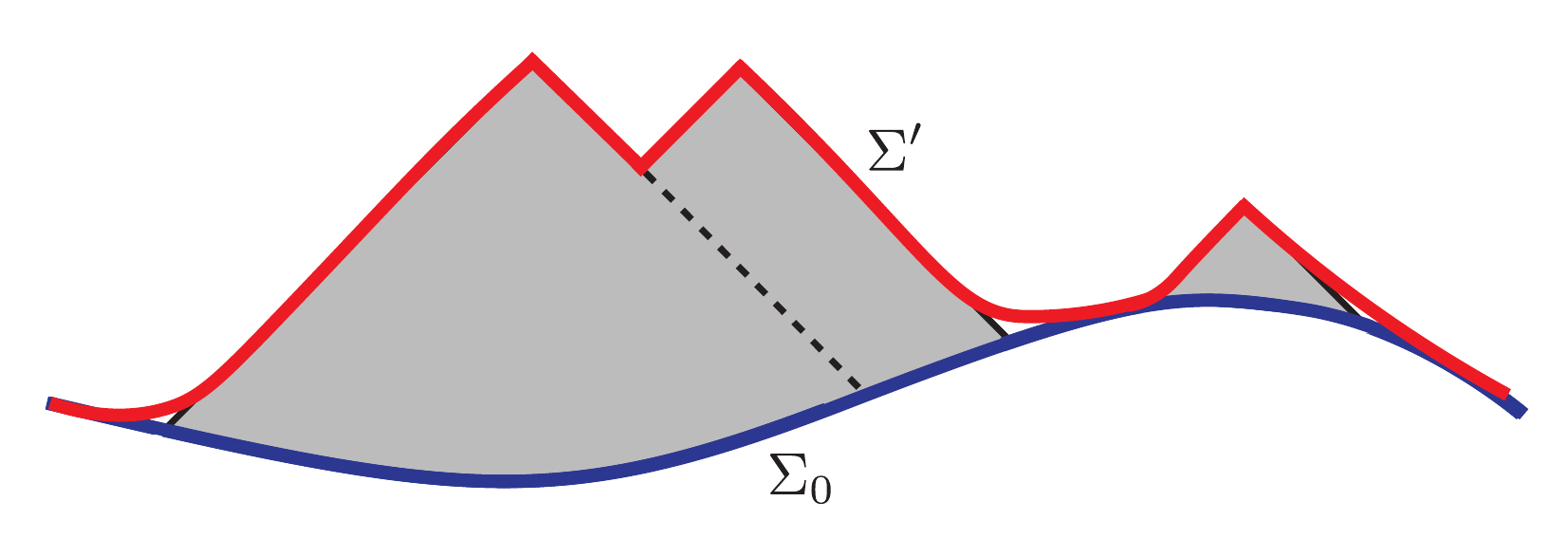}
  \caption{ The computer simulates by pushing forward the evolution pointwise, generating a new spacelike surface $\Sigma'$ (red, top) from the initial surface $\Sigma_0$ (blue, bottom). We conjecture that action integrated over the entire region between the two spacelike surfaces gives the computational cost of simulating the evolution.
    }
  \label{Simulation}
\end{figure}

Given a bulk point $p$ and an initial value surface $\Sigma_0$, we have defined an action time $t(p, \Sigma_0)$ in the multiverse as the gravitational action integrated over a certain covariantly defined region between $\Sigma_0$ and $p$: the intersection $\mathcal M = J^{-}(p)\cap I^+(\Sigma_0)$ of the causal past of $p$ with
 the future region of influence of $\Sigma_0$ (see figure~\ref{ActionTime}b). We additionally conjecture that the integrated action between two spacelike surfaces computes the computational cost of pushing the initial surface $\Sigma_0$ forward in time (see figure~\ref{Simulation}).

Including appropriate boundary terms, the action can be written as
 \begin{align} 
 S = \ & \frac{1}{16\pi G_N} \int_{\mathcal M} d^{d+1}x \sqrt{-g} \left(R-2\Lambda\right) + \frac{1}{8\pi G_N} \int_{\mathcal B} d^dx\sqrt{|h|} K \nonumber \\
 &- \frac{1}{8\pi G_N} \int_{\mathcal B'} d\lambda d^{d-1}\theta \sqrt{\gamma}\kappa + \frac{1}{8\pi G_N}\int_{\Sigma} d^{d-1}x \sqrt{\sigma} \eta + \frac{1}{8\pi G_N} \int_{\Sigma'} d^{d-1}x \sqrt{\sigma} a~.
 \end{align}
 The first term is the Einstein-Hilbert action including a cosmological constant, which is $\Lambda = d(d-1)/(2 L^2)$ in pure de Sitter with radius $ L$. The second is a Gibbons-Hawking boundary term~\cite{PhysRevD.15.2752} along spacelike or timelike boundaries $\mathcal B$ with induced metric $h_{ij}$ and extrinsic curvature $K_{ij}$. The third term is a $\kappa$ boundary term for the null boundaries $\mathcal B'$ along the past lightcone. The fourth is a Hayward joint term for the intersection of spacelike and timelike surfaces~\cite{PhysRevD.47.3275, Brill:1994mb}, and the fifth is an $a$ joint term for intersections in the case that one or more of the surfaces are null, such as along the caustic at the tip of the lightcone. The null terms have been recently worked out~\cite{Lehner:2016vdi} and studied in various examples~\cite{Chapman:2016hwi, Carmi:2016wjl, Maltz:2016max, Maltz:2016iaw} to account for subtleties of evaluating standard Gibbons-Hawking boundary terms along null surfaces.

The action time will not always make sense as a time coordinate in any spacetime, for example in pure Minkowski space where the action is identically zero. However, in the case we are concerned with, specifically an eternally inflating cosmology where the volume is dominated by positive cosmological constant vacua, we argue that it provides a sensible time coordinate in the de Sitter portions that is continuous, monotonic, and whose constant time surfaces are spacelike:\\

\begin{figure}[t!]
  \centering
  \includegraphics[width=3.5in]{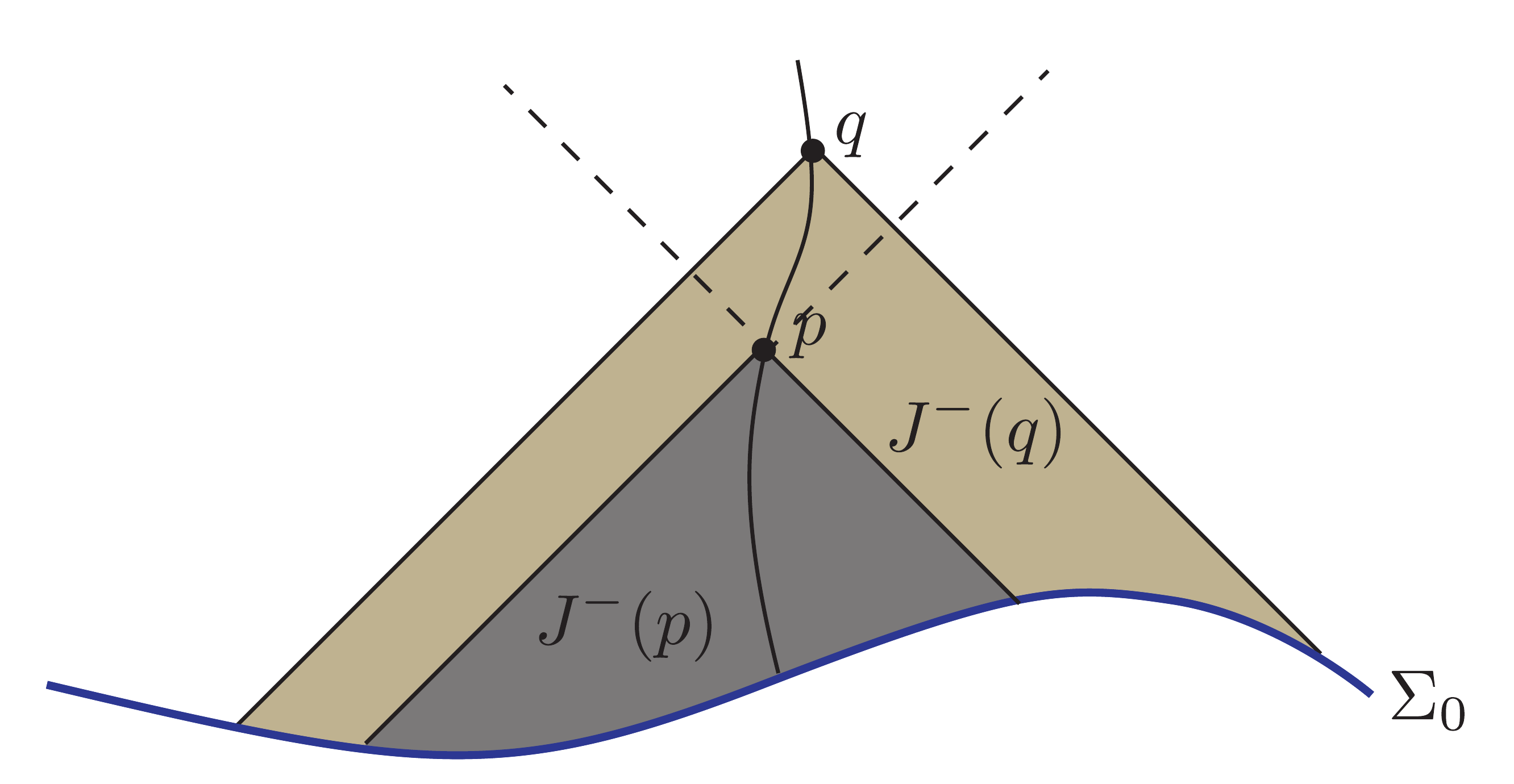}
  \caption{For action time to be well-defined, two points $p$ and $q$ that are connected along a timelike trajectory must have monotonically increasing action time. This will be the case if and only if the action integrated over $J^-(q)-J^-(p)$ (difference between light brown and dark gray regions) is always positive. A pure de Sitter vacuum with everywhere positive action trivially satisfies this requirement.}
  \label{Monotonicity}
\end{figure}

\noindent 1) {\bf Continuity}: Given an appropriate definition for the action (including boundary terms), the action has the property that it is always additive: The action of the union of two spacetime regions is the sum of the action of each individually~\cite{PhysRevD.47.3275}. From this it follows that given a spacetime point $q$ on a $4$-ball of radius $\delta$ around a point $p$, $t(q) \rightarrow t(p)$ as $\delta\rightarrow0$, in other words the action time is continuous.\\

\noindent 2) {\bf Monotonicity}: Given two timelike related spacetime points $p,q$ with $q \in J^+(p)$, the causal past of $p$ is a proper subset of the causal past of $q$: $J^-(p) \subset J^-(q)$. Thus, for fixed $\Sigma_0$ we will have $t(q) > t(p)$ so long as the action integrated over $J^-(q) - J^-(p)$ is positive. Clearly this is the case for pure de Sitter vacua, since there the action is everywhere positive. An eternally inflating cosmology will also contain regions that decay to Minkowski or anti de Sitter space, where the past lightcone of a point in de Sitter may pick up zero or negative contributions in these bubbles, but we expect that the positive de Sitter volume contribution always wins over the negative contribution from AdS regions.\\


\noindent 3) {\bf Spacelike time slices}: Constant time slices are spacelike as a consequence of monotonicity: Suppose otherwise. Then there exist some spacetime points $p, q$ with equal complexity time, $t(p) = t(q)$, where $q \in J^+(p)$. Then the past lightcone of $q$ contains as a proper subset the past lightcone of $p$, i.e. $J^-(p) \subset J^-(q)$. Assuming the complexity time is monotonically increasing, this means that $t(p) < t(q)$, a contradiction.\\

For a simple illustration, consider pure 4D de Sitter in flat slicing, with metric 
\be
ds^2 = \frac{1}{H^2} \frac{-du^2 + dx^2}{u^2}
\ee
Here $H$ is the de Sitter Hubble constant and $u<0$ is conformal time, related to proper time $\tau$ by 
$u = - (1/H) e^{-H \tau}$. Let the initial slice be at $u = a$, and consider a point $p$ at $x=0, u=b>a$. 
In the limit $b\gg a$, and neglecting boundary terms, the action of spacetime integrated over this past lightcone is 
\be
t \sim \frac{1}{G} \frac{\log(a/b)}{H^2}~,
\ee
where $G$ is the 4d Newton constant. Expressed in terms of elapsed proper time $\Delta \tau = \tau_{\rm final} - \tau_{\rm initial}$, this becomes

\be
t \sim \frac{1}{G} \frac{\Delta \tau}{H}~.
\ee

  \begin{figure}[t!]
  \centering
  \includegraphics[width=4in]{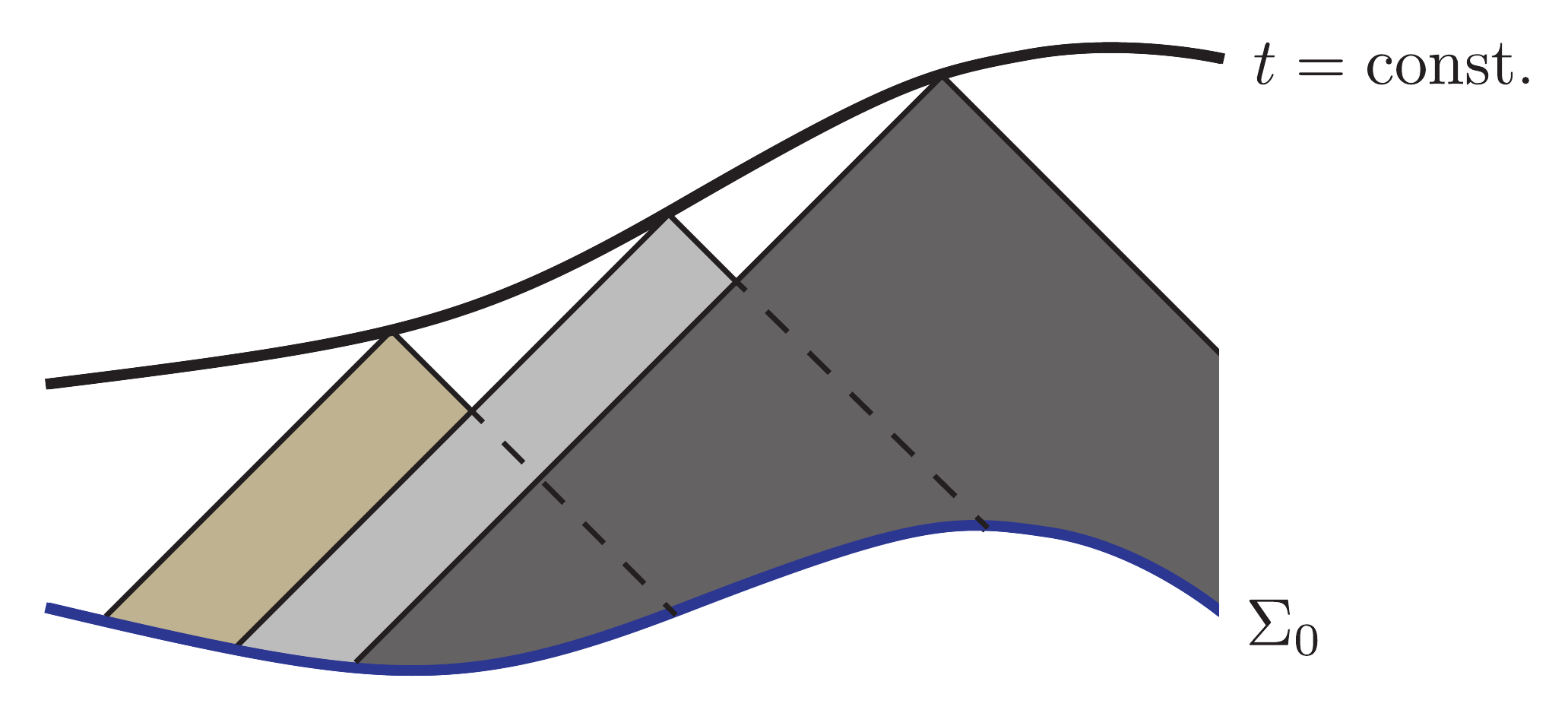}
  \caption{ A slice of constant action time in a hypothetical spacetime where the action is largest to the left of the Penrose diagram. The action integrated over each shaded region is constant. Constant action time slices are always spacelike.}
  \label{TimeEvolution}
\end{figure}

For the case of an eternally inflating universe, to zeroth order we consider an inflationary tree inflationary tree passing along a sequence of vacua $V_i$, each with Hubble constants $H_i$. The value of the action integrated across the past lightcone along this ``horizon tube" will be approximately

\be
t \sim \frac{1}{G}\sum_i \frac{\Delta \tau_i}{H_i}
\ee
where $\Delta \tau_i$ is the proper time spent in vacuum $V_i$. The total action time is the sum of contributions to the action time from each vacuum.

The action time favors \emph{large} $\Lambda$ vacua that have the smallest action time. It rewards the exponential expansion at late times even more than the proper time. Recall (see \S \ref{sec:youngness}) that the proper time had a youngness paradox due to this fact. The action time is \emph{worse} in this respect than the proper time by the same Hubble factor that, as an enhancement rather than a suppression, fixes the youngness paradox for a cutoff based on the scale factor time $\eta \sim H \tau$. Taken as a traditional global measure, the action time would have a severe youngness paradox and be ruled out. It is important that our computational measures are not standard global measures.

The search algorithm could also be equally well applied using other choices of global time coordinates, for example we already mentioned one that weights using the volume rather than the action. In fact, we expect that the results of the Markov process presented in \S \ref{sec:markov} will not be highly sensitive to the choice of time coordinate. It would be interesting to further study the sensitivity of the procedure on the choice of time coordinate.

\section{Quantum gravitational generalizations}

We have focused up to this point on a computational simulation of cosmology in the semiclassical limit. An interesting future direction concerns the extent to which our proposal generalizes to the quantum gravitational case. For now we offer a few comments about the generalization of our approach to this case.

Recall that we have defined our semiclassical measure through a cutoff procedure using a particular choice of time coordinate, the action time. In full quantum gravity, on the other hand, time is a derived concept; the state is a wave functional of 3-geometries (or 9-geometries
in string theory) which satisfies the Wheeler-de Witt equation and is thus invariant under time reparameterization.

To make contact with some of our previous definitions, suppose we say that for a given initial vacuum $init$, there is an associated wave function of the universe
$\Psi_{init}$ which solves the Wheeler-de Witt equation.  Furthermore, we grant that for each type of vacuum $i$, there is an operator
$O_i$ which is $1$ if the state contains the vacuum of type $i$ and $0$ otherwise.  The natural quantity which
measures the probability that the wave function contains vacuum $i$ is then
\be
\vev{\Psi_{init}|\; O_i\; | \Psi_{init}}~.
\ee
Thus this should be the measure factor (after normalization).

It is possible to connect with our previous discussion if we grant that in the semiclassical
regime, we have
\be \label{eq:new-HH}
\vev{\Psi_{init}|\; O_i\; | \Psi_{init}} = \sum_p e^{-t(p(i))}~,
\ee
where $t$ is the action time, in other words it is the sum of terms $e^{-S}$ over the causal past of
each $p$ which realizes the vacuum $i$.  This sum will normally be dominated by the smallest $t$
and thus the measure will be supported on the vacuum selected by the semiclassical approach.   Of course there might be degeneracies or other corrections to this result.

It would be interesting to further study the consistency of this quantum gravitational version of the proposal, for example by testing some explicit realizations of Eq.~\eqref{eq:new-HH}.

\section{Computational complexity results}
\label{s:compclass}

Having laid out in \S \ref{ss:globaltime} a set of rules obeyed by a supercomputer simulating the universe, it will now be beneficial to revisit the problem through the lens of computational complexity theory.

\subsection{Complexity theory}
There are many excellent reviews of the basics of computational complexity 
theory~\cite{DBLP:books/daglib/0072413,DBLP:books/daglib/0023084,DBLP:books/daglib/0032853,DBLP:journals/eccc/Aaronson17}, 
and in \cite{Denef:2006ad} we explain why this is relevant
for the string landscape, so we will only give a brief summary here. 
The prototypical question is: 
Given a computational problem, can we show that
a computer with {\bf limited} resources (number of elementary computations, or number of bits stored,
or perhaps conditions on both) either {\bf always can} solve the problem, or sometimes {\bf cannot}
solve the problem.  To be more precise, we consider an infinite family of related problems, 
and let the allowed resources depend on the ``size'' of the particular problem in some definite way.
The question is then, what is the minimal number of elementary computations required to 
solve the problem as a function of the problem size.

Consider for definiteness the problem of finding a factor of a positive integer $N$
(other than $1$ and $N$).
In this example, the choice of the number $N$ gives us a family of problems, and a natural definition of the
problem size is the number of digits of $N$ or equivalently its logarithm.  Thus we can ask,
what is the
minimal number of elementary computations required to factor any number $N<N_0$, 
as a function of $\log N_0$ ?
The algorithm we learn in elementary school requires listing the numbers up to $\sqrt{N}$ and
trying to divide $N$ by each one.  Granting that a division is an elementary computation, 
in the worst case where $N$ is the square of a large prime, this algorithm
will take time exponential in $\log N$.

Although there are faster algorithms than this, 
it is not known whether factoring can be done in time polynomial
in $\log N_0$.  This question is asked often enough to have its own terminology: one asks
``is factoring in \P'' ?  Why is this question more interesting than asking whether factoring
can be done in time $(\log N_0)^\alpha$ for some particular power $\alpha$, or placing
some other condition?  One reason is that more specific questions, say to find the power
$\alpha$, depend on the precise definition of elementary computation.  One could reasonably
argue that dividing two numbers of unbounded size is not elementary; one should rather 
count the number of operations on the individual digits, or Boolean gate operations.  Although
the problem and the essential nature of the algorithm is the same either way, translating an
algorithm defined using unbounded arithmetic operations to one using gate operations will
introduce additional factors of $\log N$ in the time, and perhaps change the total time.
Thus, changing the definition in this way can change the power $\alpha$, but will not change whether
these problems can be solved in time polynomial in $\log N_0$, or not.  
One says that the problem of factoring given the ability to operate on arbitrary size numbers, is reducible
in polynomial time to factoring using only Boolean gate operations.

As a more interesting example, suppose we have a Turing machine which executes programs
written on a one-dimensional tape.  As is hopefully familiar, we can
use it to solve the factoring problem by expressing our algorithm as a program on the tape,
followed by some representation of the number $N$, and running the machine.
Now the time taken by a Turing
machine just to execute a single arithmetic operation will grow as some power of $\log N$ --
without going into the details, this is because a Turing
machine has a ``head'' which sits at a particular position on its tape, and moving the head
counts as an operation.
Nevertheless, any algorithm which uses unbounded arithmetic operations can still be executed
on a Turing machine with at most a polynomial slowdown in $\log N$.  The same goes for any
class of problems solvable in polynomial time. One says that the class \P\ is 
preserved under polynomial-time reduction, so one obtains the same class of problems
under a very broad range of definitions of ``time.''

A deeper reason to be interested in \P\ is the existence of {\bf other} large and
natural classes of problems which are preserved under polynomial-time reductions.  
The most famous of these are the \NP\ problems, which are solvable in ``non-deterministic
polynomial time.''  Rather than explain the term ``non-deterministic,'' we can define these
as the problems for which the validity of a proposed solution can be {\bf checked} in polynomial
time.  Factoring is clearly in \NP, as we can easily check that a proposed factor actually divides $N$.
What is less obvious, and explained in the reviews,
is that there are \NP-complete families of problems, meaning problems for which any
problem in \NP\ can be reduced in polynomial time to one in the family.  
The main point is that the concept of computational reduction allows 
making non-trivial statements about the time required to perform a computation, which have
surprisingly little dependence on the precise definition of time.

\subsection{Complexity class of a cosmology}

Using our previous definitions we can now define the complexity class for a cosmology, which provides a natural setting to ask general questions that do not depend on the specific choice of an algorithm used by the supercomputer.

First, to match the standard definition in complexity theory, we need an infinite family of problems.
If our original problem is to find a hospitable vacuum, now we need to postulate an infinite family of 
hospitable vacua depending on a parameter.  A natural choice of parameter is to place an upper bound on
the cosmological constant, so we pose the problem \SIMCC\ to generate a hospitable vacuum
within a simulation of the multiverse,
which in addition has $|\Lambda|\le\Lambda_{max}$ for a given $\Lambda_{max}$.  We will sometimes
refer to vacua satisfying these conditions as ``target vacua'' below.

This can be compared with the problem \CC\ of \cite{Denef:2006ad}, which was to
identify the fluxes and other compactification parameters of a vacuum with $|\Lambda|\le\Lambda_{max}$
in the BP model and other toy landscape models through pure computation (not simulation).
While solving the problem \CC\ may help with solving \SIMCC, the two problems are not at all the same.
Whereas in \CC\ the computer works with mathematical representations of vacua, in \SIMCC\ it must
create them through simulation from specified initial conditions.  Thus each step in the search must
form part of a consistent history for the multiverse, and {\it a priori} this suggests that  \SIMCC\ might be harder
than \CC.

One (large) change we will make
in the present discussion is that we will take the problem size to be not $\log\Lambda_{max}$ but the quantity
$N_Q=1/\Lambda_{max}$, and thus \SIMCC\ is in \P\ if it can be solved in time $T < \Lambda_{max}^{-s}$ for
some integer $s$.  This seems natural in light of the observation that the action of an observable region for
such a universe at the time $T \sim H^{-1}$ is roughly $1/\Lambda_{max}$ and thus the complexity of simulating
this region is $N_Q$.  In these conventions, the results of \cite{Denef:2006ad} show that 
the \CC\ problem is squarely in \P\ (indeed naive search is linear).  
Thus there is no strong {\it a priori} reason to think that \SIMCC\ is not in \P.  However, a major subtlety in
this comparison is that when we posed \CC, we granted that the cosmological constant of any candidate vacuum
could be computed in fixed time, by adding and multiplying numbers in the BP model, or perhaps by solving
Picard-Fuchs equations or other mathematical problems in more realistic models.  But as was already pointed out 
in \cite{Denef:2006ad},
this claim is very dubious in actual non-supersymmetric string compactifications, 
in which the cosmological constant will receive 
contributions at all orders in perturbation theory and even non-perturbatively.  One could even entertain the
opposite conjecture, that there is no algorithm to compute the c.c. to an accuracy $\epsilon$ which is more
efficient than simulating the vacuum in a space-time region of volume $1/\epsilon$.  Still, this would only
add another factor of $N_Q$ to the total time, so given that there is such an algorithm to compute the c.c.,
the \CC\ problem could be solved in time $N_Q^2$, and based on this we
might conjecture that \SIMCC\ is in \P. 

Actually, this conjecture does not make much sense, because cosmological dynamics is probabilistic or even quantum.
Even the most unlikely tunneling events have some probability to happen quickly, so there can be no real lower
bound on the time it takes to find the target vacuum.
A more sensible question is whether \SIMCC\ is in \BPP\ or in \BQP, 
which are the probabilistic and quantum analogues of \P. 
Roughly, these classes are defined by asking that the probability of solving the problem 
in polynomial time is bounded below by a number greater than $1/2$.  Thus, for \SIMCC\ to be in
\BPP, there must be an algorithm to solve it which does not rely on unlikely events.

Another issue which must be addressed is that given an upper bound on $\Lambda$ and a lower bound on the
Kaluza-Klein mass scale $M_{KK}$, there are good arguments that the number of vacua satisfying these bounds is
finite \cite{Acharya:2006zw}, and thus there must be some minimum $\Lambda$ for vacua satisfying these bounds.
If this is not zero, we will not have an infinite family of problems.  Evidently we 
need to relax the ``quasi-realistic'' condition, by allowing compactifications with arbitrarily
small Kaluza-Klein mass scale.  One still needs to place some lower bound on the KK scale to avoid
decompactifying vacua, but perhaps this bound could depend on $\Lambda_{max}$ in a way that
both preserves the four-dimensional interpretation and allows an infinite family of problems.\footnote{
One could dispute the claim that only four-dimensional vacua can be hospitable.}

This is at least a sketch of the problem class \SIMCC, and from what we have said so far it seems not
unreasonable that it could be in \BPP\ or \BQP, indeed we might propose the following strategy to solve it.
We grant that the supercomputer can not just simulate the multiverse but that it can also work with the
equations of the fundamental theory and solve the \CC\ problem of finding the compactification
parameters of target vacua.  As we discussed, we expect that this problem can be solved in
time $N_Q^2$, to find a candidate target vacuum.  Then, we grant that the supercomputer can use the
equations to map out the landscape and compute tunneling rates.  While we are far from understanding
the landscape well enough to judge this point, at least from toy models of the sort we considered in
\S \ref{s:toymodel}, the landscape will contain paths from the initial vacuum to any other 
specified vacuum whose length is linear in the topological complexity (number of cycles etc.) of the vacuum,
and these might not be hard to find.  Then, given a path in the landscape from the
initial vacuum to a target vacuum, then one can combine the
tunneling rates to derive an expected time to traverse the path.  There is no evident need for this path to
involve very slow rates, so again we conclude that the evidence to hand is consistent with  $\mbox{\SIMCC}\in\BPP$.

This is to be compared with the estimate we made in \S \ref{s:toymodel}
of the time for the Markov process of eternal inflation to reach equilibrium, which was far longer,
controlled by the lifetime of the longest lived metastable (non-terminal) vacuum, which is a double exponential.

Let us finally discuss whether it makes sense to ask if the problem \SIMCC\ is in the class \NP.
Recall that a problem is in \NP\ if a proposed solution can be verified in time polynomial in the problem size,
possibly given additional information (a ``certificate'') also polynomial in the size.
Naively, one would say that this is the case if a vacuum with $|\Lambda|\le\Lambda_{max}$ always
arises in the multiverse (with the specified initial conditions) after an action time $T< \Lambda_{max}^{-s}$
for some $s$.  This is because we can take as the certificate the minimal history of the multiverse
which suffices to create the vacuum, in other words its past light cone intersected with the future of $\Sigma_0$.
By assumption this region can be simulated in time equal to its action, and presumably the results compared 
to the certificate in a comparable amount of time.

However, the above discussion depends on the dynamics being deterministic. In reality the dynamics is probabilistic or quantum, and this definition fails.  Suppose that as a certificate, we are given a history in which the target vacuum
is created very early through some very unlikely tunneling event.  One cannot claim that it is incorrect, just unlikely.

As we discussed earlier, the right question to ask is whether the problem of finding a viable vacuum is in \BPP\ or in \BQP.
The nondeterministic (or verification) analogs of these classes are the protocol classes \MA\ (Merlin-Arthur) and \QMA\ (the quantum version of \MA). Arthur is a computer with a random number generator which can solve polynomial time problems (in \BPP) and Merlin is an oracle with infinite computational power. Arthur is allowed to ask Merlin questions about the problem (for example, does the candidate cosmology satisfy the laws of physics), and Merlin will answer, but Arthur cannot blindly trust Merlin's answers. If there is a protocol by which Merlin can convince Arthur of the correct answer to a question with high probability, then the problem is in \MA.

To apply this to cosmology, the idea is that Merlin proposes a cosmological history in which a viable vacuum is created in polynomial time, and then Arthur checks both the equations of motion and whether any random tunneling events which took place were likely or rare by computing the amplitude using the laws of string theory. 

Furthermore, we can check whether a class of vacua $V_i$ are in \MA\ by following the time evolution along a sequence of space-like surfaces of increasing action time, and defining a probability distribution over spatial geometries where the probabilities reflect the probabilities of tunneling events between vacua. We define $\mathcal C$ to be the action time after which the probability that a vacuum in the class is created is greater than $2/3$. If $\mathcal C$ grows polynomially in $\mbox{max}_i \ \mathcal{C}_{\rm univ}(\Lambda_i)$, then the class is in \MA.

This formulation allows us to ask whether the problem of finding a given class of vacua (say de Sitter with c.c. at most $\Lambda$ is in \MA\ or \QMA. Even if it is, we can ask whether a particular way to solve the problem attains this theoretical possibility. Indeed there are many problems for which a naive algorithm is exponential, and it takes some cleverness to find a polynomial-time algorithm, for example linear programming and testing primality.

Thus, we are left with the following questions:
\begin{enumerate}[I.]
\item Is it possible to find a hospitable vacuum with $|\Lambda|\le\Lambda_{max}$
in time polynomial in $1/\Lambda_{max}$?
\item Is it possible to verify the cosmology which finds such a vacuum in polynomial time?
\item Does the usual discussion of eternal inflation find such a vacuum in polynomial time?
\item Can one at least verify such a cosmology in polynomial time?
\end{enumerate}
We argued that the answer to III and IV is no, and that the answer to II is yes. We do not know the answer to I.

\section{Conclusions}
\label{s:conclusions}

We have argued for a principle of ``limited computational complexity" that may explain why we find ourselves in our given vacuum compared to the huge number of other vacua predicted by string theory. The idea is that our vacuum should be preferentially selected as the one most easily reached by a computer simulating the evolution of the universe starting from simple initial conditions. To make this precise, we have given a prescription for obtaining a ``computational" measure factor based on this principle. This is in contrast to more traditional equilibrium methods of multiverse analysis, which necessarily evolve far into the late-time regime to compute probabilities. 

Our approach is far more dependent on the choice of initial conditions than
the equilibrium approaches, one of whose main justifications is independence
from this choice.  We argued that string theory will someday predict simple
initial conditions and that this will eventually be seen as a virtue of the proposal.
Furthermore, it does not appear to be too sensitive to fine details of  the
initial conditions, since simple compactifications can easily interconvert. 
We also gave examples illustrating the claim that the results do not much depend on details of the search algorithm used by the supercomputer.  Again, this independence is somewhat limited and we expect that among the wide variety of possible search algorithms, are choices leading to rather different results.  If so, we suggest that this choice be informed by a
stronger principle of ``minimal computational complexity,'' according to which the search algorithm with the least
expected time to find a hospitable vacuum should be preferred.  In future investigations, it would be interesting to study further the dependence on the choice of algorithm, including the precise definition of both vacua and ``hospitable vacua" as well as the restriction to minimal complexity algorithms.

The natural time coordinate used by the supercomputer is one we call action time, which we argue keeps track of computational cost of time evolution. Unlike the traditional measure program, where different choices of time coordinate result in vastly different predictions, our approach does not seem so sensitive to this choice. Nonetheless, we have argued that the action time passes several basic consistency checks necessary to use it as a clock for our simulation.

One potential issue for any measure that rewards early time evolution is that it may exhibit a ``youngness problem" by placing a huge weight on observers known as created early in the history of a universe. In our prescription this is not a consequence of the exponential growth of numbers of universes, but rather it is a result of defining the cost of the search problem to include the cost of creating observers. As such it is not solved by changing the definition of time, as in the equilibrium approach; instead it is solved by taking that cost out of the search problem, by considering a search for an anthropically hospitable universe.

An anthropically hospitable universe is not one which contains observers; rather it is one in which the physical
requisites for the creation of observers are satisfied -- semiclassical space-time,
large space-time volume, density fluctuations and structure formation, local sources of free energy such as stars, and some sort of chemistry rich enough to lead to the creation of arbitrarily complex bound states, including some which can store information.  Furthermore, it must be possible for the supercomputer to test the condition by simulating the universe, making observations and doing computations, all with a bounded definite computational cost.
Our proposal presumes that there are natural definitions of hospitableness, 
not unique definitions to be sure but for which the resulting measure does not depend much on their specifics.  
Certainly, more work should be done on this point, and we hope to return to it in future work.

Just as we avoid a youngness problem, we additionally avoid a common problem plaguing many traditional measures known as the Boltzmann brain problem. This occurs when a given measure predicts that the most likely observer is a thermal fluctuation; such observers are overwhelmingly produced at late times within a de Sitter vacuum.  Analogous
problems can happen in ``marginally hospitable''
universes on the edge of the anthropic condition, in which observers are possible granting
the occurence of extremely rare events.  Because the volumes of space-time being considered
are arbitrarily large, there is no lower bound on the rates which must be considered.

An approach that favors early time and does not postulate the existence of the entire multiverse
could circumvent these problems.  
Our proposed solution involves two elements.  First, because of the supercomputer's focus on checking for hospitable vacua rather than observers, we are no longer comparing relative probabilities of observers.  One should still worry that the check for hospitableness will occasionally be ``fooled'' by a rare event;
for example an unusually large density fluctuation.  This is possible but in our proposal the probability of such a 
rare event is not multiplied by arbitrarily large factors; rather these factors are cutoff in efficient search algorithms.
The effectiveness of this solution remains to be proven as one can imagine landscapes in which it would fail; for
example one in which the number of distinct marginally hospitable vacua far exceeds the number of ``normal''
hospitable vacua which do not rely on rare events.
But in a landscape in which numbers of vacua are roughly uniform in the parameters (c.c., $\delta\rho/\rho$, etc.)
this should not be the case.

It is worth emphasizing that with our anthropically hospitable condition we have moved away from centering observers in multiverse analysis. In this way it is fundamentally different from the standard measure paradigm. In particular, this means that certain quantities, for example the time elapsed from the big bang to our existence,
are not predictable using our computational measure, the way they would be using previous measures. The approach aims to predict the properties of the string theory vacuum we find ourselves in, along with its associated fundamental constants, but is agnostic to questions about which observer we are \emph{within} that vacuum.

Philosophically, we assert that this proposal addresses one of the most important criticisms of the multiverse.
This is the criticism that postulating a multiverse is postulating a structure which is far larger and more complex
than the observations it is being used to explain.  Our proposal directly addresses this criticism by making a
precise definition of the complexity of the universe and the multiverse -- in terms of the computational cost of
simulating each -- and makes as its founding principle the idea that the complexity of the multiverse is as small
as it could be consistent with our physical assumptions.  
It is even possible that the complexity of the multiverse is not so much larger than that of our universe.  We argued in
\S \ref{sec:ActionTime} that simulating our universe requires of order $N_Q\sim 10^{120}$ quantum gate operations.
As we discussed in \S \ref{s:compclass}, an efficient search algorithm might require
only $k\cdot N_Q$ or $k\cdot N_Q^2$ operations for some small $k$
to simulate the dynamics which led to our universe.  We do not know whether this is the case, but within 
our framework this question can actually be studied.

In \S \ref{s:measureresults} we speculated about how the computational
measure fits into the string landscape, and argued for the claim that our intuition
that there are ``simple'' and ``complicated'' string compactifications will indeed be borne out
and that the simple compactifications will turn out to be preferred as initial conditions.
Because the hospitable vacua predicted by the computational measure
tend to be as similar to the initial conditions as possible,
this leads to the prediction that the extra dimensions in our universe will have a relatively simple structure
and will realize a relatively economical way to solve the c.c. problem.
This is in contrast to the equilibrium measures, which favor vacua which can be easily reached from the
longest-lived metastable vacuum.  This vacuum is expected to be among the most complicated of vacua
\cite{Douglas:2012bu} and the vacua which can be easily reached from it are expected to be complicated as well.

This difference could easily show up in observable physics.   For example, it has been suggested that
string theory naturally favors an ``axiverse'' with hundreds or more axions \cite{Arvanitaki:2009fg}, 
because most Calabi-Yau
manifolds have hundreds of homology cycles, because these cycles are called upon to solve the c.c. problem,
and because each cycle comes with moduli which can be light.  Without claiming to study this claim
in the detail required, it is evident that a compactification with the maximal number of homology cycles
is more likely to lead to this picture than one with the minimal number required to satisfy the constraints.
We look forward to a more rigorous study of this and other potentially observable consequences of this proposal.

\section{Acknowledgements}
\label{s:acknowledgements}

This project was initiated by FD and MRD soon after the publication of \cite{Denef:2006ad}, and we thank many people for discussions over the years.
MRD would particularly like to thank Alex Vilenkin for many explanations of
quantum cosmology and measure factors, and
Raphael Bousso, Alan Guth, Andrei Linde, Dick Bond, Steve Shenker and David Kutasov
for illuminating discussions.    We also thank the referee for emphasizing the need to discuss marginally hospitable vacua.

We thank Department of Energy grant DOE DE-SC0011941 for partial support of this work. CZ is supported by NASA ATP grant NNX16AB27G. MRD thanks the organizers of the JHS 75 Meeting, the SITP Templeton Conference ``New Horizons in Inflationary Cosmology,'' and the Chris Hull Fest \cite{jhs75,sitp,hullfest} where much of the work herein was presented and discussed. As we were completing the write-up of our results, we received a paper \cite{Bao:2017thx} that touches on similar themes.

\bibliography{compinf}
\bibliographystyle{utphys}

\end{document}